\documentclass[aps,preprint,superscriptaddress]{revtex4-2}
\usepackage{graphicx}
\usepackage{txfonts}
\usepackage{bm}
\usepackage{lineno}
\usepackage{color}

\begin{document}


\title{Simultaneous manipulation of electromagnetic and elastic waves via glide symmetry phoxonic crystal waveguides}

\author{Linlin Lei}
 \affiliation{School of Physics and Materials Science, Nanchang University, Nanchang 330031, China}
\author{Lingjuan He}
 \email{helingjuan$_$123@163.com}
 \affiliation{School of Physics and Materials Science, Nanchang University, Nanchang 330031, China}
\author{Qinghua Liao}
\affiliation{School of Physics and Materials Science, Nanchang University, Nanchang 330031, China}
\author{Wenxing Liu}
\affiliation{School of Physics and Materials Science, Nanchang University, Nanchang 330031, China}
\author{Tianbao Yu}
\affiliation{School of Physics and Materials Science, Nanchang University, Nanchang 330031, China}

\date{\today}

\begin{abstract}
A phoxonic crystal waveguide with the glide symmetry is designed, in which both electromagnetic and elastic waves can propagate along the glide plane at the same time. Due to the band-sticking effect, super-cell bands of the waveguide degenerate in pairs at the boundary of the Brillouin zone, causing the appearance of gapless guided-modes in the bandgaps. The gapless guided-modes are single-modes over a relatively large frequency range. By adjusting the magnitude of the glide dislocation, the edge bandgaps of the guided-modes can be further adjusted, so as to achieve photonic and phononic single-mode guided-bands with relatively flat dispersion relationship. In addition, there exists acousto-optic interaction in the cavity constructed by the glide plane. The proposed waveguide has potential applications in the design of novel optomechanical devices.
\end{abstract}


\maketitle

\section{Introduction}
Phoxonic crystals (PXCs), in which permittivities and elastic properties are periodically arranged in the same lattice on a common wavelength scale, are designed to synchronously control the behavior of electromagnetic and elastic waves and to enhance the opto-acoustic interaction\cite{Maldovan2006,Maldovan2006a,Eichenfield2009,Favero2009,Rolland2012,Lucklum2013,Rolland2014,Escalante2014,Pennec2010,Mohammadi2010,
Kipfstuhl2014,Ma2017,Moradi2018,Qiu2020,Jin2021}. The most significant characteristic of PXCs is the simultaneous photonic and phononic bandgaps, or the phoxonic bandgaps, resulting from multiple scattering of photons and phonons, in which transmission of waves is forbidden\cite{Aram2018}. The existence of phoxonic bandgaps provides an opportunity for designing phoxonic devices and functional materials\cite{Xia2019,Aboutalebi2021,Ma2022,Lei2022}. Past researches have shown that by introducing a linear gapped defect into an otherwise perfect PXC, electromagnetic and elastic waves can travel along the defect at the same time when guided modes are excited, forming a PXC gapped waveguide\cite{Lin2013,Chiu2017,Ma2014}. The waveguide not only can guide the waves but also can enhance their interaction because of the simultaneous confinement of photons and phonons\cite{Laude2011,Lin2013,Shu2020}. However, the waveguides depending on the gapped defects are usually multimodal, leading a competition of the guided bands inside the complete bandgap that can severely flatten the guided bands\cite{Martinez2022}. Thus, it is necessary to explore new operation degree of freedom to achieve single-mode phoxonic guided modes. 

Decades ago, structures with higher symmetries, including twist, glide and their combination, have been utilized to design novel waveguides\cite{QUEVEDOTERUEL2021}. However, only recently has there been a resurgence of interest in these higher symmetries in order to manipulate electromagnetic and airborne acoustic waves. A periodic structure has a glide symmetry (GS) if it remains invariant after a translation and a mirroring with respect to a plane called the glide plane. The most salient characteristics the GS given are the band-sticking or degeneracy at the boundary of the first Brillouin zone (FBZ), which offers a new route to design the band dispersion and gives rise to distinctive applications in kinds of waveguides. For example, GS can be used to decrease the dispersion and to widen the bandwidth\cite{Dahlberg2017}. Furthermore, GS can be able to improve gain and bandwidth of leaky-wave and lens antennas, and can boost the performance of phase shifters and filters realized in standard and groove-gap waveguide technologies\cite{Beadle2019,Jankovic2021,Abdollahy2021}. These properties were found in various periodic structures, confirming that the benefits of GS are applicable to a wide range of practical waveguide devices. These waveguides are usually one-dimension (1D) periodic structures with only one unit cell perpendicular to the direction of the waveguide. The GS waveguides can also be extended to the two-dimensional (2D) periodic structures, like 2D photonic and phononic crystals\cite{Mahmoodian2016,Yoshimi2020,Yoshimi2021,Xie2021,Martinez2022,Wen2022}. For instance, GS combined with a linear gapped defect can be used to regulate the dispersion relationship of guided modes in a triangular photonic crystal\cite{Mock2010}. GS waveguide without extra gapped linear defect has been realized in a square lattice phononic crystal to obtain gapless and GS-protected acoustic guided-modes\cite{Martinez2022}. 

In this letter, we introduce the GS into PXCs, which enables simultaneous guidance of electromagnetic and elasitc waves along the glide plane. The GS provides an alternative way to to manipulate electromagnetic and elastic waves at the same time. Moreover, the glide parameter that quantifies the magnitude of the dislocation can be used to adjust the size of the edge bandgaps, offering the waveguide a function of filtering electromagnetic and elasitc waves at the same time. The glide interface can also be used to construct PXC cavities to enhance the acousto-optic (AO) interaction. Although the realization of the PXC waveguide is based on a square lattice, the principle can also be extended to other lattice types, like triangular lattices and honeycomb lattices.
\section{Geometry and band structures of the PXC}
\begin{figure*}[ht!]
	\centering
		\includegraphics[scale=0.2]{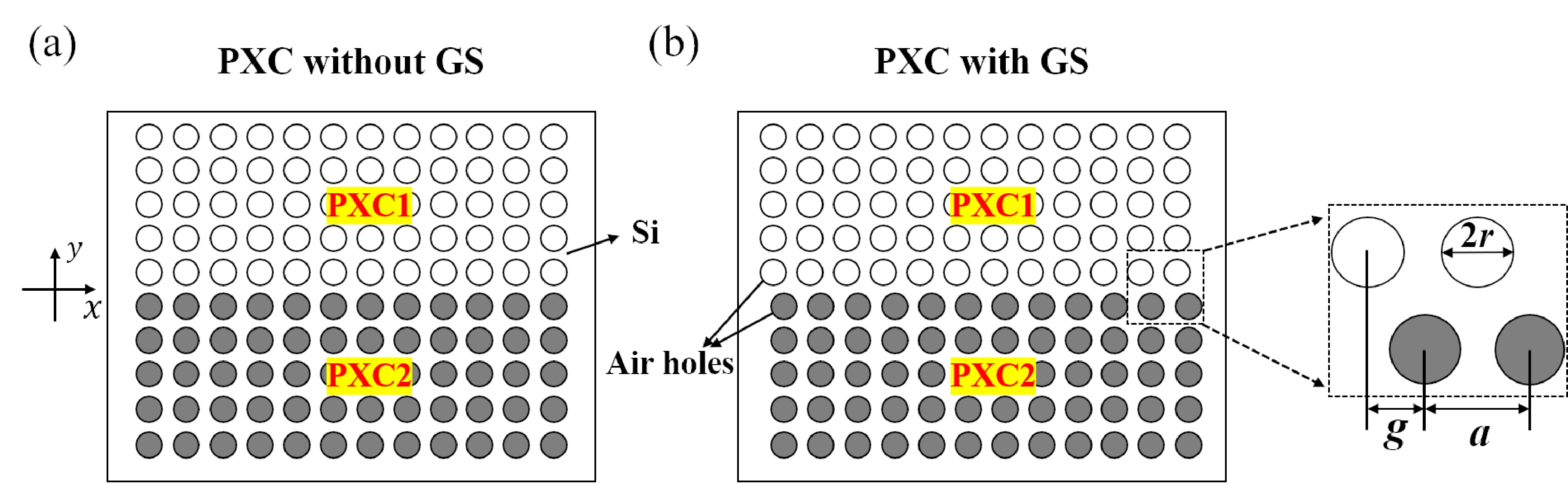}
	\caption{Schematic of the two-dimensional (2D) PXC waveguide with air holes embedded into a Si host, spliced by two pieces of identical PXCs with air holes colored by white and grey, respectively. The two identical PXCs are labeled as PXC1 and PXC2. (a) PXC with glide parameter $g=0$. (b) PXC with glide parameter $g=a/2$. The enlarged view in (b) shows the glide parameter $g$, lattice constant $a$, and diameter $2r$ of air holes.}
	\label{fig_1}
\end{figure*}
The 2D PXC is made of silicon with air holes periodically arranged in a square lattice, as shown in Fig.1(a). The lattice constant $a$ is set to be 400 $nm$ and the radius $r$ of the air hole is $0.46a$. Introducing a glide dislocation with a non-zero glide parameter $g$ along the glide plane $y=0$, one half of the PXC is shifted by a certain distance in the $x$ direction. The two sub-crystals are labeled in Fig. 1 as PXC1 and PXC2. Although the resultant PXC in Fig.1(b) losses a translational symmetry in the $y$ direction but gains the GS when $g=a/2$\cite{Kim2020,Ghasemifard2018,Zhang2017,Nica2015}. A glide symmetric structure overlaps with itself after a translation of $a/2$ and a reflection with respect to the glide plane. The material parameters of silicon are as follows: mass density $\rho = 2331~kg/m^3$, relative permittivity $\varepsilon = 12.5$, transverse and longitudinal sound speed $c_t = 5360~m/s$ and $c_l = 8950~m/s$, respectively\cite{Lei2021,Lei2022}. The three independent stiffness coefficients of the silion material are: $C_{11}=16.57\times10^{10} Nm^{-21}$, $C_{12}=6.39\times10^{10} Nm^{-21}$,  and $C_{44}=7.962\times10^{10} Nm^{-21}$\cite{Shu2020a}. Herein, the calculation of band structures and numerical simulations are based on the finite element methods.
\begin{figure*}[ht!]
	\centering
		\includegraphics[scale=0.09]{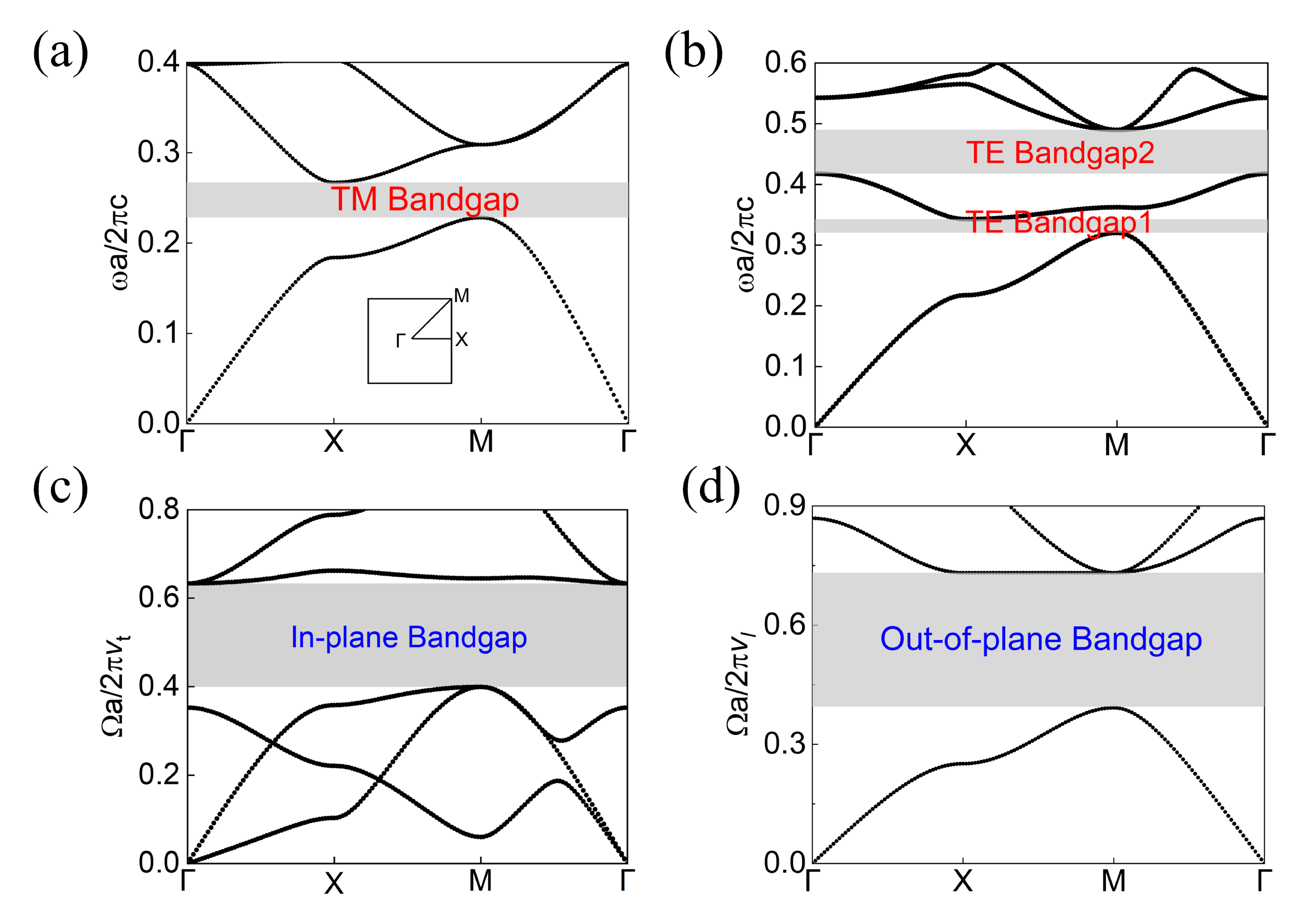}
	\caption{Band structures of the PXC unit-cell. (a) Transverse magnetic (TM) modes having a complete bandgap between the first and second bands. (b) Transverse electric (TE) modes having two complete bandgaps, one between the first and second bands and the other between the second and third bands. (b) In-plane and (d) out-plane elastic modes, both of which show one complete bandgap.}
	\label{fig_2}
\end{figure*}

We plot the phoxonic band structures for the unit-cell using Floquet periodic boundary conditions, as shown in Fig. 2. For photonic modes, we consider transverse magnetic (TM) modes with electric fields along the axis of the air holes and transverse electric (TE) modes with magnetic fields along the axis of the air holes. The TM and TE band structures are shown in Figs. 2(a) and 2(b), showing one and two complete bandgaps, respectivley. Inset in Fig.2(a) is the first Brillouin zone (FBZ), showing the highest symmetry points. For phononic modes, Figs. 2(c) and 2(d) show the band structures of in-plane and out-of-plane elastic modes, respectively, both of which possess one complete bandgap. We would like to clarify two points here. Although normalized center frequencies of photonic and phononic bandgaps are in the order of $10^{-1}$, but the actual phononic frequency is typically orders of magnitude smaller than the photonic frequency for a common lattice constant $a$ since the sound speed is much smaller than the light speed. Besides, their is no GR for the case in Fig. 1(a), and the unit-cell bands do not show the pairwise degenerate points at the X point of the FBZ.
\begin{figure*}[ht!]
	\centering
		\includegraphics[scale=0.09]{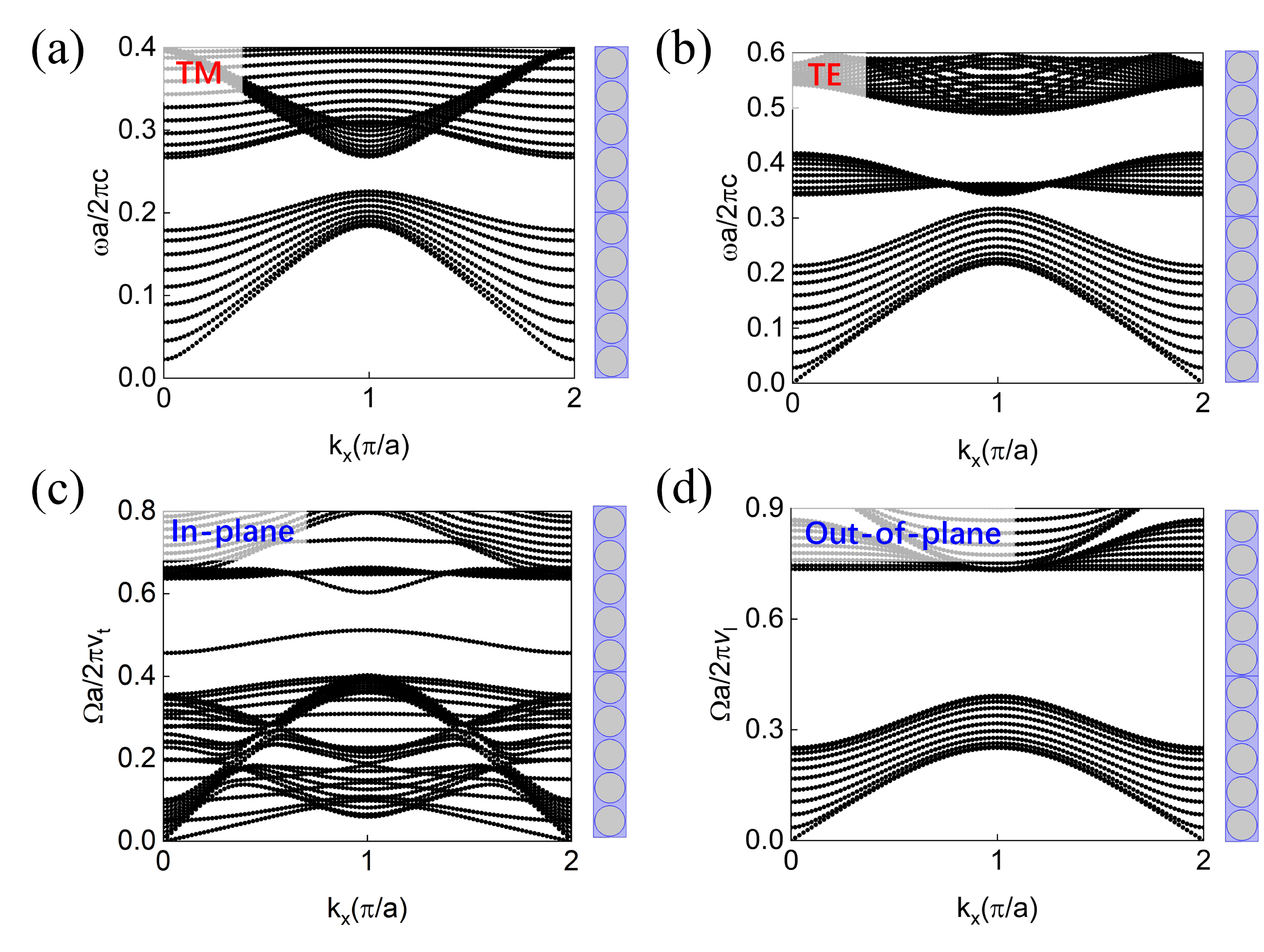}
	\caption{Band structures of the PXC super-cell with $g=0$ for (a) TM, (b) TE, (c) in-plane and (d) out-plane elastic modes. There are no guided modes lying in the super-cell bandgaps.}
	\label{fig_3}
\end{figure*}
\begin{figure*}[ht!]
	\centering
		\includegraphics[scale=0.09]{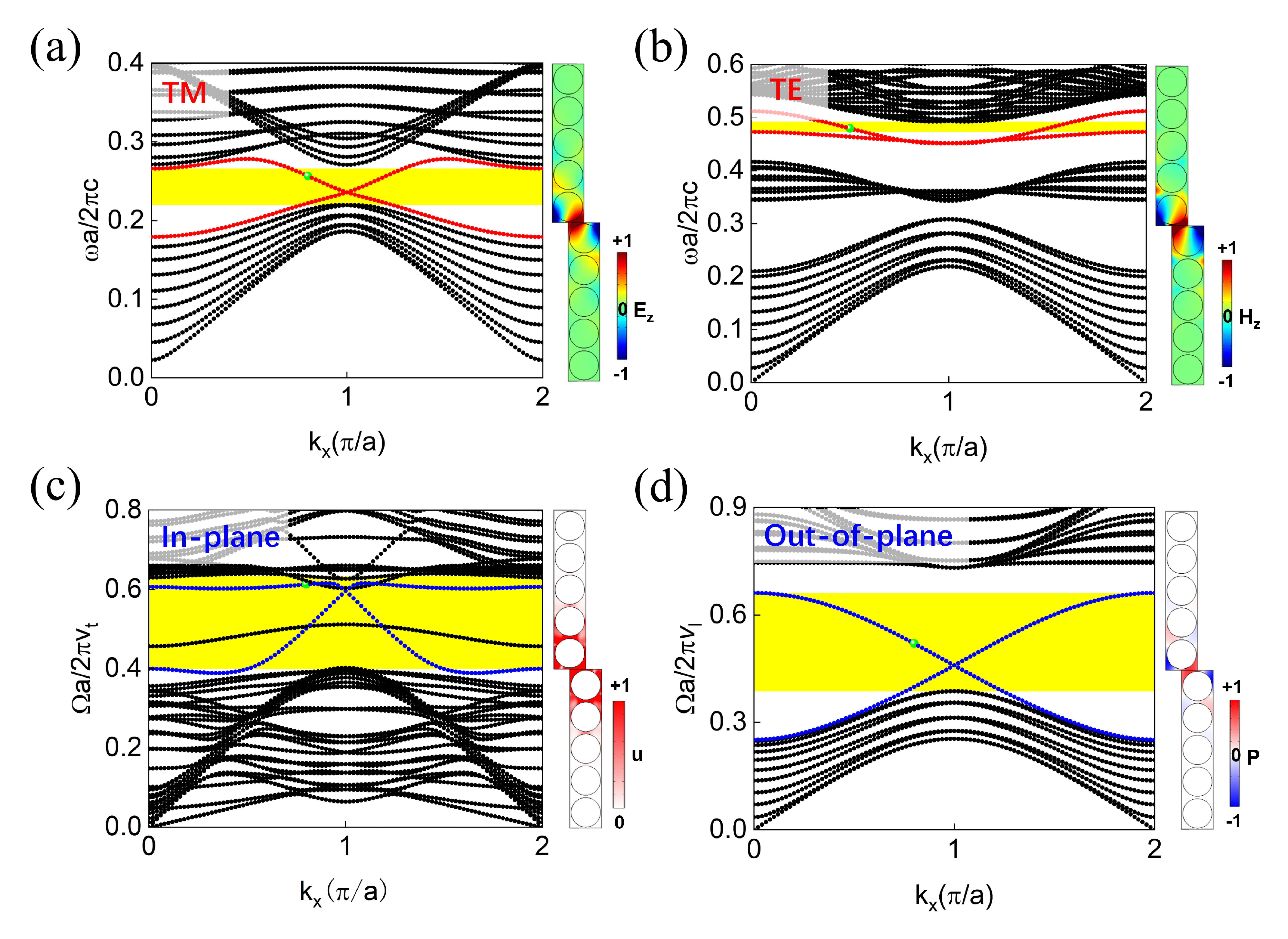}
	\caption{Band structures of the PXC super-cell with $g=a/2$ for (a) TM, (b) TE, (c) In-plane and (d) out-plane elastic modes. All the bands have to pairwise degenerate at $k_x=\pi/a$, the boundary of the FBZ of the super-cell, and there are two gapless guided modes located in the super-cell bandgaps. The green dot in each panel is one of the eigenmodes of the guided bands, and its eigenfield is shown in the right of the panel.}
	\label{fig_4}
\end{figure*}

\begin{figure*}[ht!]
	\centering
		\includegraphics[scale=0.1]{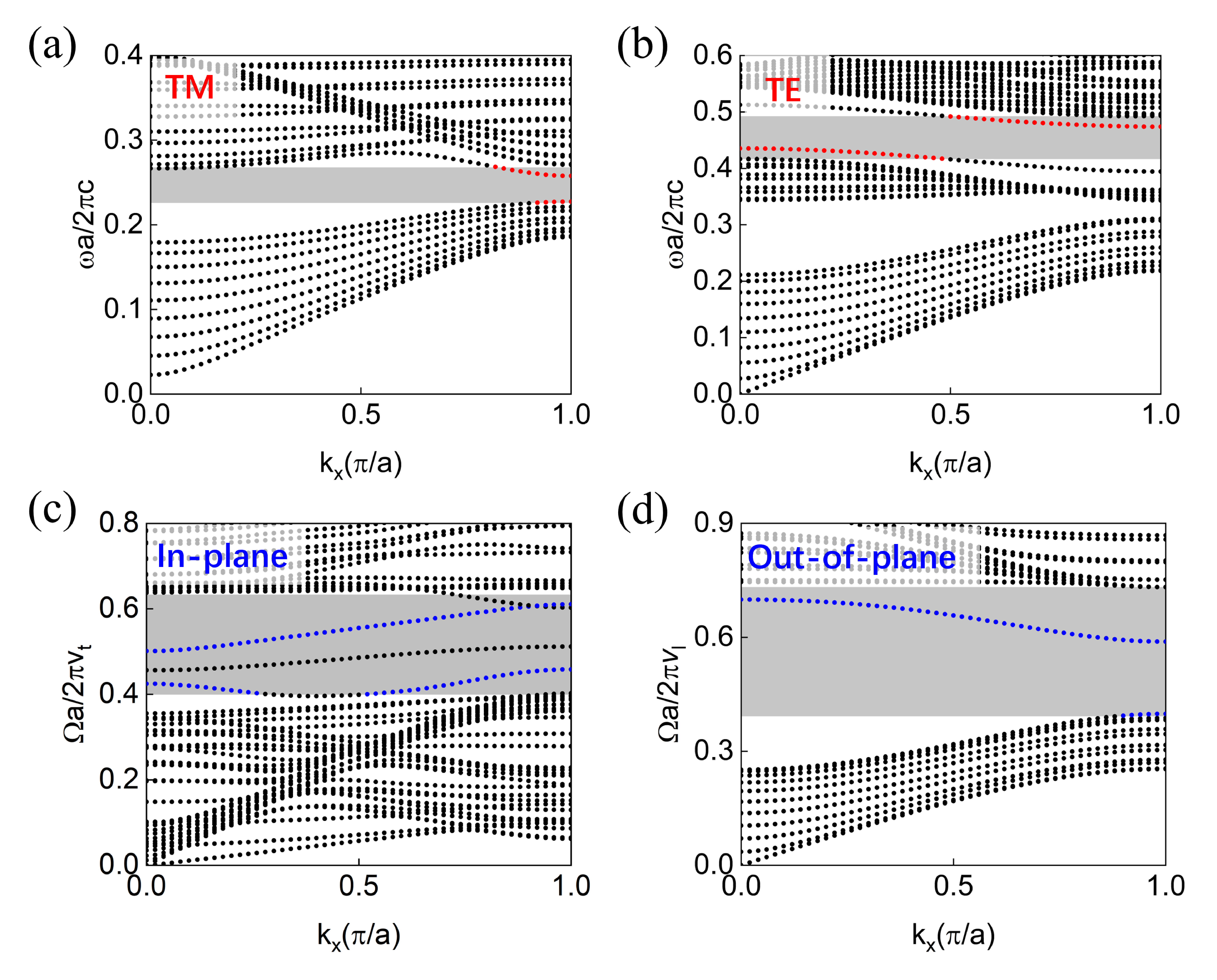}
	\caption{Band structures of the PXC super-cell with $g=a/4$ for (a) TM, (b) TE, (c) In-plane and (d) out-plane elastic modes. Band degeneracies vanish, and edge bandgaps appear at the boundary of the FBZ of the super-cell. The shaded areas denote the bandgaps.}
	\label{fig_5}
\end{figure*}  
\begin{figure*}[ht!]
	\centering  
		\includegraphics[scale=0.2]{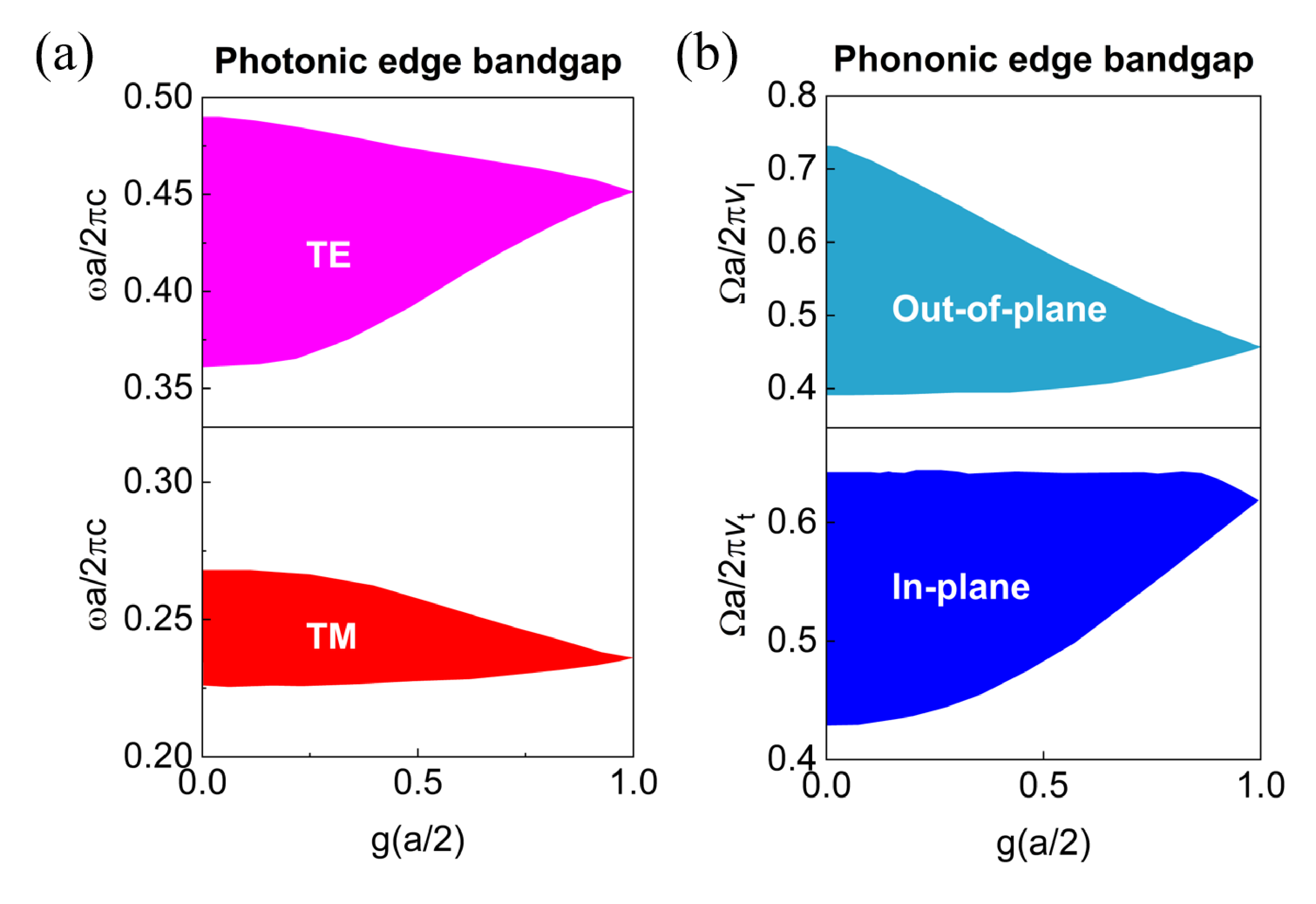}
	\caption{Evolution processes of (a) photonic and (b) phononic edge bandgaps. Red and pink areas denote the bandgaps of TM and TE modes, while blue and cyan areas denote that of the in-plane and out-of-plane elastic modes, respectively. As $g$ goes from 0 to $a/2$, the phoxonic edge bandgaps gradually close to zero.}
	\label{fig_6}
\end{figure*}

\section{Regulation of the glide dislocation on the phoxonic guided modes}
To explore the existence of the phoxonic guided modes, we plot the band structures of the supercell composed of 10 unit-cells with $g=0$, as shown in Fig.3. In the calculation, periodic boundary conditions are applied to the $x$ direction, while perfect-electric-conductor and free boundary conditions are applied to the $y$ direction for the photonic and phononic modes, respectively\cite{Martinez2022,Lei2023}. Not surprisingly, there are no guided modes located in the photonic (Figs. 3(a) and 3(b)) and phononic (Figs. 3(c) and 3(d)) bandgaps due to the absence of linear gapped defects or the topological phase transition along the $y$ direction between the PXC1 and PXC2. Of note that the doubly degenerated bands located in the in-plane elastic bandgap are the surface waves that travel along the upper and lower boundaries of the PXC made of the supercells\cite{Martinez2022}. 

However, this situation will be changed when the GR is introduced into the PXC. The GR is achieved by a translation of the PXC2 with $g=a/2$. As a result, pairwise degenerate points would appear at the boundary of the FBZ of the super-cell. To verify this, super-cell band structures for TM, TE, in-plane and out-of-plane elastic modes are plotted in Figs. 4(a)-4(d), respectively, from which we can see that all the bands are degenerate at the $k_x=\pi/a$. Moreover, there are two gapless modes located in the phoxonic bandgap, where there should be nothing. The right panels of Figs. 4(a), 4(c), and 4(d) show the fields of $E_z$, total displace $u$, and pressure P for TM, in-plane, and out of plane in-gap modes for the higher-frequency branches at $k_x=0.8(\pi/a)$, while the right panel of Fig. 4(b) shows the $H_z$ field of TE in-gap modes at $k_x=0.5(\pi/a)$, respectively. The positions of these modes are labelled by the green dots on the in-gap phoxonic modes. As can be seen, these phoxonic in-gapped modes are well confined at the interface between the PXC1 and PXC2, and the same thing is true for the lower-frequency branches of the in-gap modes. Therefore, they are interface modes not the surface modes that travel along the upper and lower boundaries of the PXC. Moreover, the photonic and phononic guided modes are single-mode within a relatively large frequency range. As the yellow shaded areas shown in Fig. 4, the frequency ranges of the single-modes are $0.21954(2\pi c/a)\sim 0.26590(2\pi c/a)$, $0.47274(2\pi c/a)\sim 0.49228(2\pi c/a)$, $0.43890(2\pi v_t/a)\sim 0.60028(2\pi v_t/a)$, and $0.38695(2\pi v_l/a)\sim 0.66134(2\pi v_l/a)$ for TM, TE, in-plane and out-of-plane elastic modes, respectively. 

If the glide parameter $g$ deviates $a/2$, the gapless in-gap modes will be changed into gapped modes due to the broken of the GS. Fig. 5 shows the super-cell bands for the case of $g=a/4$, and the shaded areas denote the bandgaps. As can be seen, all the band degeneracies vanish, and the edge bandgap that is defined as the bandgap of the two in-gap modes at the boundary of the FBZ appears. As a result, there are single-mode photonic and phononic guided modes located in their respective bandgaps, with relatively flat dispersion relationship. In particular, the group velocities of the in-gap modes approach to zero near the center and the boundary of FBZ. This would help to enhance the interaction between the electromagnetic waves and elastic waves due to the prolonged interaction time\cite{Yu2018}. It is worth noting that the gapless guided-modes could only exit for $g=a/2$. Fig. 6 gives a visualized process of the phoxonic guided-mode degeneracies. As $g$ goes from 0 to $a/2$, the phoxonic edge bandgaps gradually close to zero at $a/2$. Thus, glide dislocation provides a degree of freedom to manipulate the phoxonic dispersion relationship and the size of edge bandgaps.  

Of note, not all the bulk bandgaps have the in-gap modes. In Fig.4(b), the in-gap TE modes only exit in the second super-cell bandgap. In order to provide an explanation for the band degeneracy and why there are in-gap gapless modes, we introduce the glide symmetry operator $\hat{G}$, which is defined as $\hat{G}\psi(x,y)=\psi(x+a/2,-y)$ , where $\psi$ is the Bloch wave function. 
The combination of the glide symmetry operator $\hat{G}$ and time-reversal operator $\hat{\theta}$ gives $\hat{\Theta}=\hat{G}\hat{\theta}$. When acting on the $\psi$ twice, $\hat{\Theta}^2\psi=e^{ik_xa}\psi$. Thus, $\hat{\Theta}^2\psi=-\psi$ at $k_x=\pi/a$, which forms the Kramers-like degeneracy 
and makes all bands to group in pairs at the $k_x=\pi/a$ including the nearest two bands on either side of the phoxonic band gaps\cite{Lin2020}. Thus, if there is an odd number of bands below the super-cell bandgap, two in-gap gapless modes could exist; if there is an even number of bands below the super-cell bandgap, no in-gap gapless modes could exist since there is no extra band to degenerate with the bands above the bandgap. There are ten bands below the first TE super-cell bandgap, and thus no in-gap modes could exist there.  

\section{PXC waveguide with GS}
As the phoxonic in-gap modes can be confined at the glide plane with nonvanishing group velocities, waves hence can travel along the glide plane when operating frequencies lie in the frequency range of in-gap modes. To verify this, a waveguide made of the PXC with GS is constructed, and Figs. 7(a)-7(d) shows field distributions of $E_z$, $H_z$, in-plane and out-of-plane elastic modes when electromagnetic and elastic waves are incident from the left. The frequencies of the waves for TM, TE, in-plane, and out-of-plane modes are $0.25733(c/a)$, $0.48000(c/a)$, $0.56716(v_t/a)$, and $0.52737(v_l/a)$, respectively. As can be seen, the waves only travel along the glide plane with little energy permeated into the bulk PXC. Thus, these phoxonic in-gap modes are confined guided-modes. 
\begin{figure*}[ht!]
	\centering
		\includegraphics[scale=0.18]{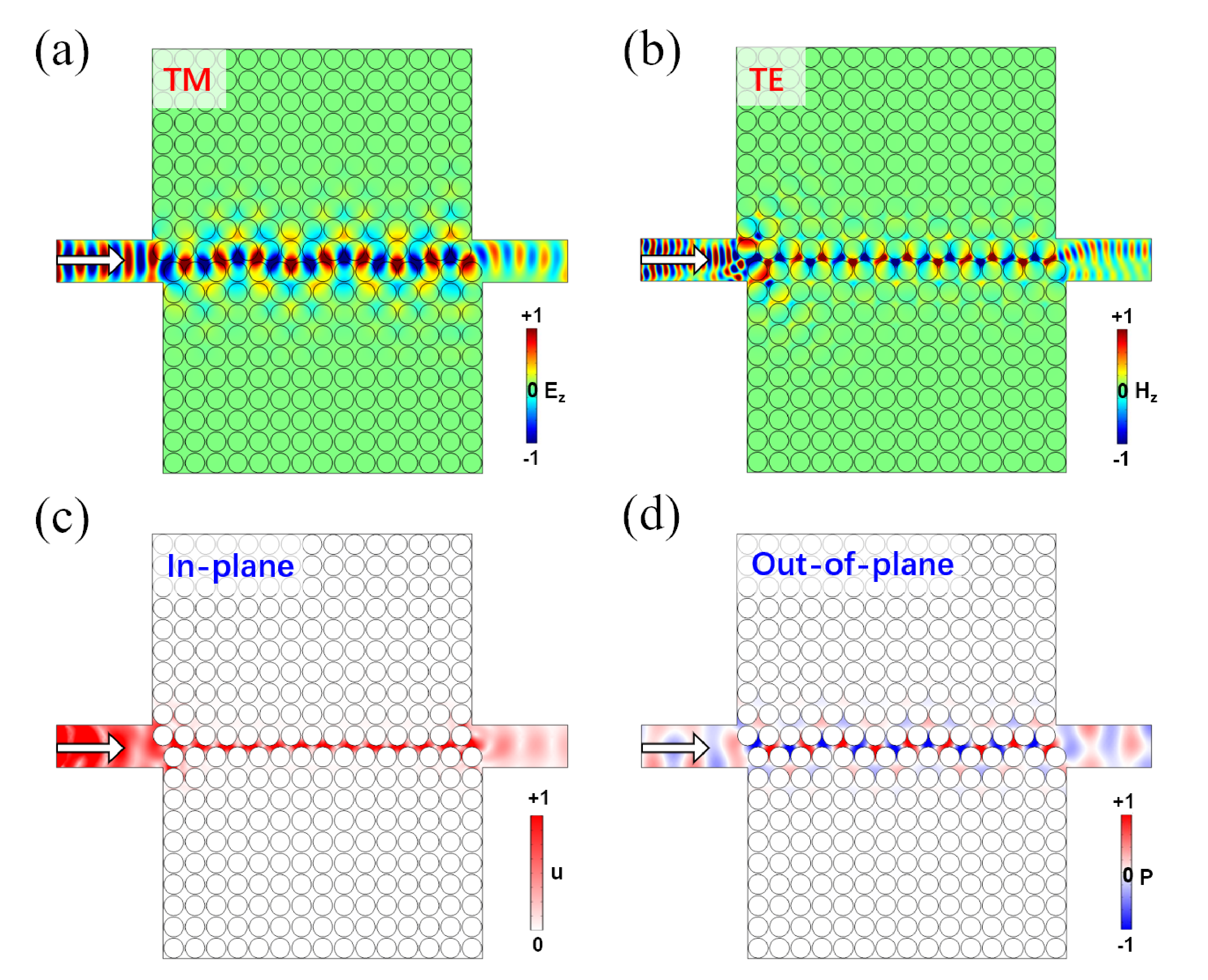}
	\caption{Field distributions of the phoxonic waveguide with GS for (a) TM, (b) TE, (c) in-plane, and (d) out-of-plane modes with normalized frequencies $0.25733(c/a)$, $0.48000(c/a)$, $0.56716(v_t/a)$, and $0.52737(v_l/a)$, respectively, from which we can see that electromagnetic and elastic waves can only travel along the glide plane. The waves are incident from the left, indicated by the white arrows.}
	\label{fig_7}
\end{figure*}   
\begin{figure*}[ht!]
	\centering
		\includegraphics[scale=0.18]{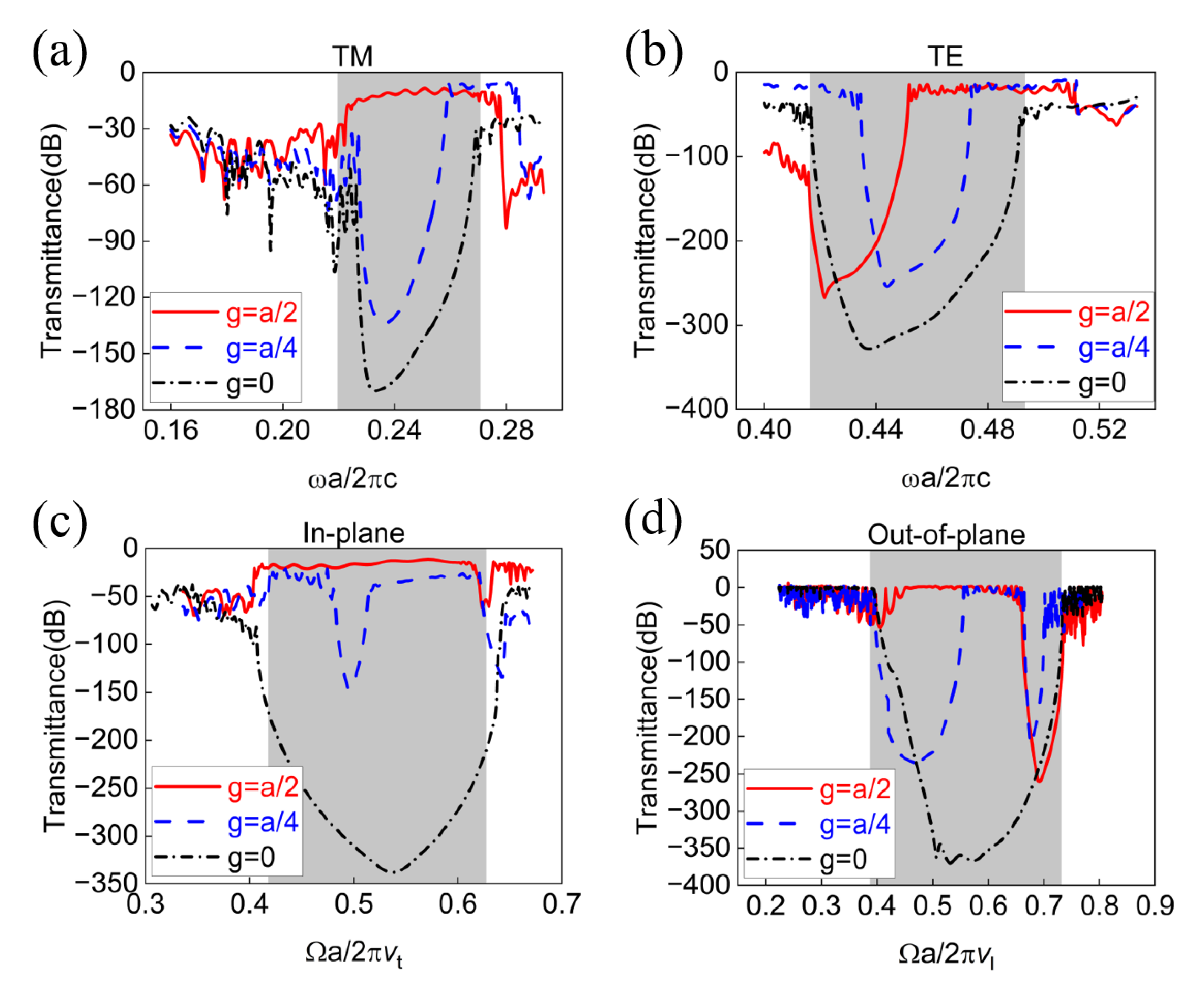}
	\caption{Transmittance of (a) TM, (b) TE, (c) in-plane, and (d) out-of-plane elastic modes. Red solid lines, blue dashed lines, and black dash-and-dot lines denote transmittance of the waveguide for $g=a/2$, $g=a/4$, and $g=0$, respectively. Shaded areas denote super-cell bandgaps.}
	\label{fig_8}
\end{figure*}   

We further plot the transmittance for waveguides with and without GS in Fig. 8, indicated by the red solid lines and black dot-and-dash lines, respectively. For the PXC with GS, due to the existance of the gapless guided-modes, both the photonic (Figs. 8(a) and 8(b)) and phononic (Figs. 8(c) and 8(d)) modes show continuous and high transmittance within the super-cell bandgaps (indicated by the dashed areas). The blue dashed-lines in Fig. 8 show the phoxonic transmittance for $g=a/4$. Compared with the transmittance for $g=a/2$, extra transmittance dips appear for the TM, in-plane, and out-of-plane modes, while the transmittance dip moves towards the high-frequency range for TE modes. This is because when $g=a/4$ the phoxonic edge bandgaps are open, and the waves that used to pass through the GS waveguide cannot now. For the PXC with $g=0$, no wave can enter the waveguide, since no guided modes exist in the bandgaps, as the black dot-dashed lines shown in Fig. 8. Thus, the glide parameter $g$ offers a controllable variable to adjust the in-gap modes to change from gapless to gapped guided-modes for both the photonic and phononic modes, which contributes to select desired waveguide frequencies and to eliminate unwanted ones. 

\section{Acousto-optic (AO) interaction}
\begin{figure*}[ht!]
	\centering
		\includegraphics[scale=0.10]{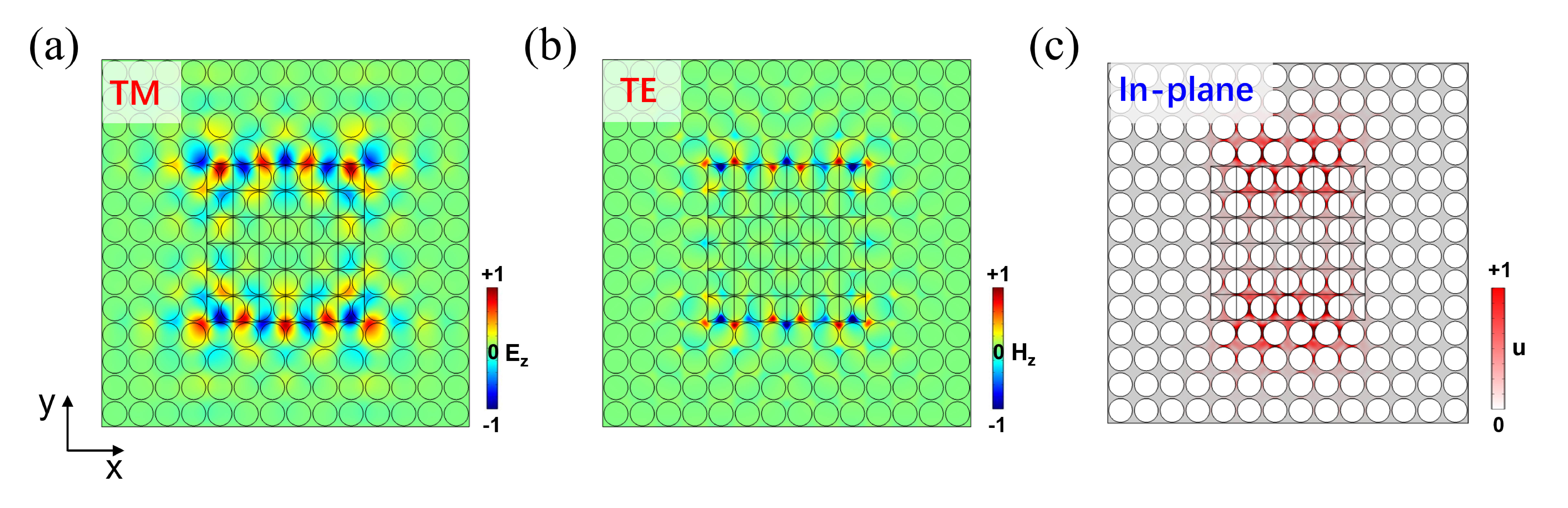}
	\caption{Eigenfields of the cavity modes for TM, TE, and in-plane elastic modes, and their normalized frequencies are 0.25908(c/a), 0.48604(c/a), and 0.60263($v_t$/a), respectively.}
	\label{fig_9}
\end{figure*}   
Here, we use the glide interface to construct a PXC cavity, as shown in Fig. 9. As can be seen, the electromagnetic waves and in-plane elastic waves can be well confined into the upper and lower boundaries of the inner boundary at the same time, and their normalized frequencies are 0.25908(c/a), 0.48604(c/a), and 0.60263($v_t$/a), respectively. The photonic and phononic eigenfield distribution are symmetric about the $y$-axis. The simultaneous confinement would enhance the AO interaction, in which the moving interfaces (MIs) and photoelastic (PE) effects are considered. The MI effect is due to the dynamic motion of the silicon–air interfaces, while the PE effect is related to the change of the refractive index by the generation of the strain field in the structure\cite{Shu2020a}. The full optomechanical coupling rate is $g_{\text{OM}}=g_{\text{OM,MI}}+g_{\text{OM,PE}}$. The PE contribution by unit thickness of the PXC is\cite{Eljallal2013}
\begin{equation}
g_{\text{OM,PE}}=-\frac{\omega_0}{2}\frac{\langle E|\frac{\partial\varepsilon}{\partial\alpha}|E\rangle}{\int \boldsymbol{E}^*\cdot D dS}.
\end{equation}
where
\begin{equation}
\langle E|\frac{\partial\varepsilon}{\partial\alpha}|E\rangle=-\epsilon_0 n^4\int|E_2|^2(p_{12}S_{xx}+p_{12}S_{yy})dS,
\end{equation}
for the TM modes, and
\begin{equation}
\langle E|\frac{\partial\varepsilon}{\partial\alpha}|E\rangle=-\epsilon_0 n^4\int[2Re{E_x^*E_y}p_{44}S_{xy}+|E_x|^2(p_{11}S_xx+p_12S_{yy})+|E_y|^2(p_{11}S_{yy}+p_{12}S_{xx})]dS
\end{equation}
for the TE modes.

The MI contribution by unit thickness of the PXC is given by
\begin{equation}
g_{\text{OM,MI}}=-\frac{\omega_0}{2}\frac{\int (\boldsymbol{U}\cdot\boldsymbol{n})(\Delta\varepsilon\cdot\boldsymbol{E}_\parallel^2-\Delta\varepsilon^{-1}\cdot\boldsymbol{D_\perp}^2)dl)}{\int \boldsymbol{E}^*\cdot D dS}.
\end{equation}
where $\boldsymbol{U}$ is the normalized displacement field, $\boldsymbol{n}$ is the outward facing surface normal, $\boldsymbol{E}$ is the electric field, $\boldsymbol{D}$ is the displacement field. The subscripts $\parallel$ and $\perp$ indicate respectively the component parallel and perpendicular to the surface of the electromagnetic field. The PE coefficients of silicon are: $p_{11}=-0.1$, $p_{12}=0.01$, and $p_{44}=-0.05$. $\Delta\varepsilon=\varepsilon_{material}-\varepsilon_{air}$ and $\Delta\varepsilon^{-1}\equiv\varepsilon_{material}^{-1}-\varepsilon_{air}^{-1}$, where $\varepsilon$ is the material permittivity. The integration is performed along the interfaces between the circular contour line of the hole and the Si matrix boundaries.
The calculated optomechanical coupling rates per $1nm$ thickness of the PXC for TM and TE modes are $g_{\text{OM}}=g_{\text{OM,MI}}+g_{\text{OM,PE}}=-0.000022(c/a)-0.000001(c/a)=-0.000023(c/a)$ and $g_{OM}=g_{OM,MI}+g_{OM,PE}=0.000015(c/a)-0.000003(c/a)=0.000012(c/a)$, respectively. The nonzero coupling rates show there exists interaction between photons and phonons, which would provide potential application in optomechanical devices.
\section{Conclusion}
In summary, a PXC waveguide with GS is proposed. Due to the band-sticking effect, a pair of gapless guided-modes appear in the phoxonic bandgaps. Furthermore, by changing the magnitude of the glide dislocation, the edge bandgaps and the dispersion relationship of the guided modes can be further adjusted, which helps to simultaneously achieve photonic and phononic single-mode guided bands with relatively flat dispersion relationship. By  constructing a PXC cavity, there exists AO interaction due to the simultaneous confinement of electromagnetic waves and in-plane elastic waves. Our work has potential applications in the design of optomechanical devices, and the principle of realization can also be extended to other lattice types, like triangular lattices and honeycomb lattices.  

\bibliography{refs}

\begin{thebibliography}{47}%
\makeatletter
\providecommand \@ifxundefined [1]{%
 \@ifx{#1\undefined}
}%
\providecommand \@ifnum [1]{%
 \ifnum #1\expandafter \@firstoftwo
 \else \expandafter \@secondoftwo
 \fi
}%
\providecommand \@ifx [1]{%
 \ifx #1\expandafter \@firstoftwo
 \else \expandafter \@secondoftwo
 \fi
}%
\providecommand \natexlab [1]{#1}%
\providecommand \enquote  [1]{``#1''}%
\providecommand \bibnamefont  [1]{#1}%
\providecommand \bibfnamefont [1]{#1}%
\providecommand \citenamefont [1]{#1}%
\providecommand \href@noop [0]{\@secondoftwo}%
\providecommand \href [0]{\begingroup \@sanitize@url \@href}%
\providecommand \@href[1]{\@@startlink{#1}\@@href}%
\providecommand \@@href[1]{\endgroup#1\@@endlink}%
\providecommand \@sanitize@url [0]{\catcode `\\12\catcode `\$12\catcode
  `\&12\catcode `\#12\catcode `\^12\catcode `\_12\catcode `\%12\relax}%
\providecommand \@@startlink[1]{}%
\providecommand \@@endlink[0]{}%
\providecommand \url  [0]{\begingroup\@sanitize@url \@url }%
\providecommand \@url [1]{\endgroup\@href {#1}{\urlprefix }}%
\providecommand \urlprefix  [0]{URL }%
\providecommand \Eprint [0]{\href }%
\providecommand \doibase [0]{https://doi.org/}%
\providecommand \selectlanguage [0]{\@gobble}%
\providecommand \bibinfo  [0]{\@secondoftwo}%
\providecommand \bibfield  [0]{\@secondoftwo}%
\providecommand \translation [1]{[#1]}%
\providecommand \BibitemOpen [0]{}%
\providecommand \bibitemStop [0]{}%
\providecommand \bibitemNoStop [0]{.\EOS\space}%
\providecommand \EOS [0]{\spacefactor3000\relax}%
\providecommand \BibitemShut  [1]{\csname bibitem#1\endcsname}%
\let\auto@bib@innerbib\@empty
\bibitem [{\citenamefont {Maldovan}\ and\ \citenamefont
  {Thomas}(2006{\natexlab{a}})}]{Maldovan2006}%
  \BibitemOpen
  \bibfield  {author} {\bibinfo {author} {\bibfnamefont {M.}~\bibnamefont
  {Maldovan}}\ and\ \bibinfo {author} {\bibfnamefont {E.}~\bibnamefont
  {Thomas}},\ }\bibfield  {title} {\bibinfo {title} {Simultaneous complete
  elastic and electromagnetic band gaps in periodic structures},\ }\href
  {https://doi.org/10.1007/s00340-006-2241-y} {\bibfield  {journal} {\bibinfo
  {journal} {Applied Physics B}\ }\textbf {\bibinfo {volume} {83}},\ \bibinfo
  {pages} {595} (\bibinfo {year} {2006}{\natexlab{a}})}\BibitemShut {NoStop}%
\bibitem [{\citenamefont {Maldovan}\ and\ \citenamefont
  {Thomas}(2006{\natexlab{b}})}]{Maldovan2006a}%
  \BibitemOpen
  \bibfield  {author} {\bibinfo {author} {\bibfnamefont {M.}~\bibnamefont
  {Maldovan}}\ and\ \bibinfo {author} {\bibfnamefont {E.~L.}\ \bibnamefont
  {Thomas}},\ }\bibfield  {title} {\bibinfo {title} {Simultaneous localization
  of photons and phonons in two-dimensional periodic structures},\ }\href
  {https://doi.org/10.1063/1.2216885} {\bibfield  {journal} {\bibinfo
  {journal} {Applied Physics Letters}\ }\textbf {\bibinfo {volume} {88}},\
  \bibinfo {pages} {251907} (\bibinfo {year} {2006}{\natexlab{b}})}\BibitemShut
  {NoStop}%
\bibitem [{\citenamefont {Eichenfield}\ \emph {et~al.}(2009)\citenamefont
  {Eichenfield}, \citenamefont {Camacho}, \citenamefont {Chan}, \citenamefont
  {Vahala},\ and\ \citenamefont {Painter}}]{Eichenfield2009}%
  \BibitemOpen
  \bibfield  {author} {\bibinfo {author} {\bibfnamefont {M.}~\bibnamefont
  {Eichenfield}}, \bibinfo {author} {\bibfnamefont {R.}~\bibnamefont
  {Camacho}}, \bibinfo {author} {\bibfnamefont {J.}~\bibnamefont {Chan}},
  \bibinfo {author} {\bibfnamefont {K.~J.}\ \bibnamefont {Vahala}},\ and\
  \bibinfo {author} {\bibfnamefont {O.}~\bibnamefont {Painter}},\ }\bibfield
  {title} {\bibinfo {title} {A picogram- and nanometre-scale photonic-crystal
  optomechanical cavity},\ }\href {https://doi.org/10.1038/nature08061}
  {\bibfield  {journal} {\bibinfo  {journal} {Nature}\ }\textbf {\bibinfo
  {volume} {459}},\ \bibinfo {pages} {550} (\bibinfo {year}
  {2009})}\BibitemShut {NoStop}%
\bibitem [{\citenamefont {Favero}\ and\ \citenamefont
  {Karrai}(2009)}]{Favero2009}%
  \BibitemOpen
  \bibfield  {author} {\bibinfo {author} {\bibfnamefont {I.}~\bibnamefont
  {Favero}}\ and\ \bibinfo {author} {\bibfnamefont {K.}~\bibnamefont
  {Karrai}},\ }\bibfield  {title} {\bibinfo {title} {Optomechanics of
  deformable optical cavities},\ }\href
  {https://doi.org/10.1038/nphoton.2009.42} {\bibfield  {journal} {\bibinfo
  {journal} {Nature Photonics}\ }\textbf {\bibinfo {volume} {3}},\ \bibinfo
  {pages} {201} (\bibinfo {year} {2009})}\BibitemShut {NoStop}%
\bibitem [{\citenamefont {Rolland}\ \emph {et~al.}(2012)\citenamefont
  {Rolland}, \citenamefont {Oudich}, \citenamefont {El-Jallal}, \citenamefont
  {Dupont}, \citenamefont {Pennec}, \citenamefont {Gazalet}, \citenamefont
  {Kastelik}, \citenamefont {L{\'{e}}v{\^{e}}que},\ and\ \citenamefont
  {Djafari-Rouhani}}]{Rolland2012}%
  \BibitemOpen
  \bibfield  {author} {\bibinfo {author} {\bibfnamefont {Q.}~\bibnamefont
  {Rolland}}, \bibinfo {author} {\bibfnamefont {M.}~\bibnamefont {Oudich}},
  \bibinfo {author} {\bibfnamefont {S.}~\bibnamefont {El-Jallal}}, \bibinfo
  {author} {\bibfnamefont {S.}~\bibnamefont {Dupont}}, \bibinfo {author}
  {\bibfnamefont {Y.}~\bibnamefont {Pennec}}, \bibinfo {author} {\bibfnamefont
  {J.}~\bibnamefont {Gazalet}}, \bibinfo {author} {\bibfnamefont {J.~C.}\
  \bibnamefont {Kastelik}}, \bibinfo {author} {\bibfnamefont {G.}~\bibnamefont
  {L{\'{e}}v{\^{e}}que}},\ and\ \bibinfo {author} {\bibfnamefont
  {B.}~\bibnamefont {Djafari-Rouhani}},\ }\bibfield  {title} {\bibinfo {title}
  {Acousto-optic couplings in two-dimensional phoxonic crystal cavities},\
  }\href {https://doi.org/10.1063/1.4744539} {\bibfield  {journal} {\bibinfo
  {journal} {Applied Physics Letters}\ }\textbf {\bibinfo {volume} {101}},\
  \bibinfo {pages} {061109} (\bibinfo {year} {2012})}\BibitemShut {NoStop}%
\bibitem [{\citenamefont {Lucklum}\ \emph {et~al.}(2013)\citenamefont
  {Lucklum}, \citenamefont {Zubtsov},\ and\ \citenamefont
  {Oseev}}]{Lucklum2013}%
  \BibitemOpen
  \bibfield  {author} {\bibinfo {author} {\bibfnamefont {R.}~\bibnamefont
  {Lucklum}}, \bibinfo {author} {\bibfnamefont {M.}~\bibnamefont {Zubtsov}},\
  and\ \bibinfo {author} {\bibfnamefont {A.}~\bibnamefont {Oseev}},\ }\bibfield
   {title} {\bibinfo {title} {Phoxonic crystals{\textemdash}a new platform for
  chemical and biochemical sensors},\ }\href
  {https://doi.org/10.1007/s00216-013-7093-9} {\bibfield  {journal} {\bibinfo
  {journal} {Analytical and Bioanalytical Chemistry}\ }\textbf {\bibinfo
  {volume} {405}},\ \bibinfo {pages} {6497} (\bibinfo {year}
  {2013})}\BibitemShut {NoStop}%
\bibitem [{\citenamefont {Rolland}\ \emph {et~al.}(2014)\citenamefont
  {Rolland}, \citenamefont {Dupont}, \citenamefont {Gazalet}, \citenamefont
  {Kastelik}, \citenamefont {Pennec}, \citenamefont {Djafari-Rouhani},\ and\
  \citenamefont {Laude}}]{Rolland2014}%
  \BibitemOpen
  \bibfield  {author} {\bibinfo {author} {\bibfnamefont {Q.}~\bibnamefont
  {Rolland}}, \bibinfo {author} {\bibfnamefont {S.}~\bibnamefont {Dupont}},
  \bibinfo {author} {\bibfnamefont {J.}~\bibnamefont {Gazalet}}, \bibinfo
  {author} {\bibfnamefont {J.-C.}\ \bibnamefont {Kastelik}}, \bibinfo {author}
  {\bibfnamefont {Y.}~\bibnamefont {Pennec}}, \bibinfo {author} {\bibfnamefont
  {B.}~\bibnamefont {Djafari-Rouhani}},\ and\ \bibinfo {author} {\bibfnamefont
  {V.}~\bibnamefont {Laude}},\ }\bibfield  {title} {\bibinfo {title}
  {Simultaneous bandgaps in {LiNbO}3 phoxonic crystal slab},\ }\href
  {https://doi.org/10.1364/oe.22.016288} {\bibfield  {journal} {\bibinfo
  {journal} {Optics Express}\ }\textbf {\bibinfo {volume} {22}},\ \bibinfo
  {pages} {16288} (\bibinfo {year} {2014})}\BibitemShut {NoStop}%
\bibitem [{\citenamefont {Escalante}\ \emph {et~al.}(2014)\citenamefont
  {Escalante}, \citenamefont {Mart{\'{\i}}nez},\ and\ \citenamefont
  {Laude}}]{Escalante2014}%
  \BibitemOpen
  \bibfield  {author} {\bibinfo {author} {\bibfnamefont {J.~M.}\ \bibnamefont
  {Escalante}}, \bibinfo {author} {\bibfnamefont {A.}~\bibnamefont
  {Mart{\'{\i}}nez}},\ and\ \bibinfo {author} {\bibfnamefont {V.}~\bibnamefont
  {Laude}},\ }\bibfield  {title} {\bibinfo {title} {Design of single-mode
  waveguides for enhanced light-sound interaction in honeycomb-lattice silicon
  slabs},\ }\href {https://doi.org/10.1063/1.4864661} {\bibfield  {journal}
  {\bibinfo  {journal} {Journal of Applied Physics}\ }\textbf {\bibinfo
  {volume} {115}},\ \bibinfo {pages} {064302} (\bibinfo {year}
  {2014})}\BibitemShut {NoStop}%
\bibitem [{\citenamefont {Pennec}\ \emph {et~al.}(2010)\citenamefont {Pennec},
  \citenamefont {Rouhani}, \citenamefont {Boudouti}, \citenamefont {Li},
  \citenamefont {Hassouani}, \citenamefont {Vasseur}, \citenamefont
  {Papanikolaou}, \citenamefont {Benchabane}, \citenamefont {Laude},\ and\
  \citenamefont {Martinez}}]{Pennec2010}%
  \BibitemOpen
  \bibfield  {author} {\bibinfo {author} {\bibfnamefont {Y.}~\bibnamefont
  {Pennec}}, \bibinfo {author} {\bibfnamefont {B.~D.}\ \bibnamefont {Rouhani}},
  \bibinfo {author} {\bibfnamefont {E.~H.~E.}\ \bibnamefont {Boudouti}},
  \bibinfo {author} {\bibfnamefont {C.}~\bibnamefont {Li}}, \bibinfo {author}
  {\bibfnamefont {Y.~E.}\ \bibnamefont {Hassouani}}, \bibinfo {author}
  {\bibfnamefont {J.~O.}\ \bibnamefont {Vasseur}}, \bibinfo {author}
  {\bibfnamefont {N.}~\bibnamefont {Papanikolaou}}, \bibinfo {author}
  {\bibfnamefont {S.}~\bibnamefont {Benchabane}}, \bibinfo {author}
  {\bibfnamefont {V.}~\bibnamefont {Laude}},\ and\ \bibinfo {author}
  {\bibfnamefont {A.}~\bibnamefont {Martinez}},\ }\bibfield  {title} {\bibinfo
  {title} {Simultaneous existence of phononic and photonic band gaps in
  periodic crystal slabs},\ }\href {https://doi.org/10.1364/OE.18.014301}
  {\bibfield  {journal} {\bibinfo  {journal} {Opt. Express}\ }\textbf {\bibinfo
  {volume} {18}},\ \bibinfo {pages} {14301} (\bibinfo {year}
  {2010})}\BibitemShut {NoStop}%
\bibitem [{\citenamefont {Mohammadi}\ \emph {et~al.}(2010)\citenamefont
  {Mohammadi}, \citenamefont {Eftekhar}, \citenamefont {Khelif},\ and\
  \citenamefont {Adibi}}]{Mohammadi2010}%
  \BibitemOpen
  \bibfield  {author} {\bibinfo {author} {\bibfnamefont {S.}~\bibnamefont
  {Mohammadi}}, \bibinfo {author} {\bibfnamefont {A.~A.}\ \bibnamefont
  {Eftekhar}}, \bibinfo {author} {\bibfnamefont {A.}~\bibnamefont {Khelif}},\
  and\ \bibinfo {author} {\bibfnamefont {A.}~\bibnamefont {Adibi}},\ }\bibfield
   {title} {\bibinfo {title} {Simultaneous two-dimensional phononic and
  photonic band gaps in opto-mechanical crystal slabs},\ }\href
  {https://doi.org/10.1364/OE.18.009164} {\bibfield  {journal} {\bibinfo
  {journal} {Opt. Express}\ }\textbf {\bibinfo {volume} {18}},\ \bibinfo
  {pages} {9164} (\bibinfo {year} {2010})}\BibitemShut {NoStop}%
\bibitem [{\citenamefont {Kipfstuhl}\ \emph {et~al.}(2014)\citenamefont
  {Kipfstuhl}, \citenamefont {Guldner}, \citenamefont {Riedrich-Möller},\ and\
  \citenamefont {Becher}}]{Kipfstuhl2014}%
  \BibitemOpen
  \bibfield  {author} {\bibinfo {author} {\bibfnamefont {L.}~\bibnamefont
  {Kipfstuhl}}, \bibinfo {author} {\bibfnamefont {F.}~\bibnamefont {Guldner}},
  \bibinfo {author} {\bibfnamefont {J.}~\bibnamefont {Riedrich-Möller}},\ and\
  \bibinfo {author} {\bibfnamefont {C.}~\bibnamefont {Becher}},\ }\bibfield
  {title} {\bibinfo {title} {Modeling of optomechanical coupling in a phoxonic
  crystal cavity in diamond},\ }\href {https://doi.org/10.1364/oe.22.012410}
  {\bibfield  {journal} {\bibinfo  {journal} {Optics Express}\ }\textbf
  {\bibinfo {volume} {22}},\ \bibinfo {pages} {12410} (\bibinfo {year}
  {2014})}\BibitemShut {NoStop}%
\bibitem [{\citenamefont {Ma}\ \emph {et~al.}(2017)\citenamefont {Ma},
  \citenamefont {Wang},\ and\ \citenamefont {Zhang}}]{Ma2017}%
  \BibitemOpen
  \bibfield  {author} {\bibinfo {author} {\bibfnamefont {T.-X.}\ \bibnamefont
  {Ma}}, \bibinfo {author} {\bibfnamefont {Y.-S.}\ \bibnamefont {Wang}},\ and\
  \bibinfo {author} {\bibfnamefont {C.}~\bibnamefont {Zhang}},\ }\bibfield
  {title} {\bibinfo {title} {Simultaneous guidance of surface acoustic and
  surface optical waves in phoxonic crystal slabs},\ }\href
  {https://doi.org/10.3390/cryst7110350} {\bibfield  {journal} {\bibinfo
  {journal} {Crystals}\ }\textbf {\bibinfo {volume} {7}},\ \bibinfo {pages}
  {350} (\bibinfo {year} {2017})}\BibitemShut {NoStop}%
\bibitem [{\citenamefont {Moradi}\ and\ \citenamefont
  {Bahrami}(2018)}]{Moradi2018}%
  \BibitemOpen
  \bibfield  {author} {\bibinfo {author} {\bibfnamefont {P.}~\bibnamefont
  {Moradi}}\ and\ \bibinfo {author} {\bibfnamefont {A.}~\bibnamefont
  {Bahrami}},\ }\bibfield  {title} {\bibinfo {title} {Design of an
  optomechanical filter based on solid/solid phoxonic crystals},\ }\href
  {https://doi.org/10.1063/1.5018840} {\bibfield  {journal} {\bibinfo
  {journal} {Journal of Applied Physics}\ }\textbf {\bibinfo {volume} {123}},\
  \bibinfo {pages} {115113} (\bibinfo {year} {2018})}\BibitemShut {NoStop}%
\bibitem [{\citenamefont {Qiu}\ \emph {et~al.}(2020)\citenamefont {Qiu},
  \citenamefont {Shomroni}, \citenamefont {Seidler},\ and\ \citenamefont
  {Kippenberg}}]{Qiu2020}%
  \BibitemOpen
  \bibfield  {author} {\bibinfo {author} {\bibfnamefont {L.}~\bibnamefont
  {Qiu}}, \bibinfo {author} {\bibfnamefont {I.}~\bibnamefont {Shomroni}},
  \bibinfo {author} {\bibfnamefont {P.}~\bibnamefont {Seidler}},\ and\ \bibinfo
  {author} {\bibfnamefont {T.~J.}\ \bibnamefont {Kippenberg}},\ }\bibfield
  {title} {\bibinfo {title} {Laser cooling of a nanomechanical oscillator to
  its zero-point energy},\ }\href
  {https://doi.org/10.1103/physrevlett.124.173601} {\bibfield  {journal}
  {\bibinfo  {journal} {Physical Review Letters}\ }\textbf {\bibinfo {volume}
  {124}},\ \bibinfo {pages} {173601} (\bibinfo {year} {2020})}\BibitemShut
  {NoStop}%
\bibitem [{\citenamefont {Jin}\ \emph {et~al.}(2021)\citenamefont {Jin},
  \citenamefont {Wang}, \citenamefont {Zhan},\ and\ \citenamefont
  {Hu}}]{Jin2021}%
  \BibitemOpen
  \bibfield  {author} {\bibinfo {author} {\bibfnamefont {J.}~\bibnamefont
  {Jin}}, \bibinfo {author} {\bibfnamefont {X.}~\bibnamefont {Wang}}, \bibinfo
  {author} {\bibfnamefont {L.}~\bibnamefont {Zhan}},\ and\ \bibinfo {author}
  {\bibfnamefont {H.}~\bibnamefont {Hu}},\ }\bibfield  {title} {\bibinfo
  {title} {Strong quadratic acousto-optic coupling in 1d multilayer phoxonic
  crystal cavity},\ }\href {https://doi.org/10.1515/ntrev-2021-0034} {\bibfield
   {journal} {\bibinfo  {journal} {Nanotechnology Reviews}\ }\textbf {\bibinfo
  {volume} {10}},\ \bibinfo {pages} {443} (\bibinfo {year} {2021})}\BibitemShut
  {NoStop}%
\bibitem [{\citenamefont {Aram}\ and\ \citenamefont
  {Khorasani}(2018)}]{Aram2018}%
  \BibitemOpen
  \bibfield  {author} {\bibinfo {author} {\bibfnamefont {M.~H.}\ \bibnamefont
  {Aram}}\ and\ \bibinfo {author} {\bibfnamefont {S.}~\bibnamefont
  {Khorasani}},\ }\bibfield  {title} {\bibinfo {title} {Optomechanical coupling
  strength in various triangular phoxonic crystal slab cavities},\ }\href
  {https://doi.org/10.1364/josab.35.001390} {\bibfield  {journal} {\bibinfo
  {journal} {Journal of the Optical Society of America B}\ }\textbf {\bibinfo
  {volume} {35}},\ \bibinfo {pages} {1390} (\bibinfo {year}
  {2018})}\BibitemShut {NoStop}%
\bibitem [{\citenamefont {Xia}\ \emph {et~al.}(2019)\citenamefont {Xia},
  \citenamefont {Fan},\ and\ \citenamefont {Liu}}]{Xia2019}%
  \BibitemOpen
  \bibfield  {author} {\bibinfo {author} {\bibfnamefont {B.}~\bibnamefont
  {Xia}}, \bibinfo {author} {\bibfnamefont {H.}~\bibnamefont {Fan}},\ and\
  \bibinfo {author} {\bibfnamefont {T.}~\bibnamefont {Liu}},\ }\bibfield
  {title} {\bibinfo {title} {Topologically protected edge states of phoxonic
  crystals},\ }\href {https://doi.org/10.1016/j.ijmecsci.2019.02.037}
  {\bibfield  {journal} {\bibinfo  {journal} {International Journal of
  Mechanical Sciences}\ }\textbf {\bibinfo {volume} {155}},\ \bibinfo {pages}
  {197} (\bibinfo {year} {2019})}\BibitemShut {NoStop}%
\bibitem [{\citenamefont {Aboutalebi}\ and\ \citenamefont
  {Bahrami}(2021)}]{Aboutalebi2021}%
  \BibitemOpen
  \bibfield  {author} {\bibinfo {author} {\bibfnamefont {S.~Z.}\ \bibnamefont
  {Aboutalebi}}\ and\ \bibinfo {author} {\bibfnamefont {A.}~\bibnamefont
  {Bahrami}},\ }\bibfield  {title} {\bibinfo {title} {Design of phoxonic filter
  using locally-resonant cavities},\ }\href
  {https://doi.org/10.1088/1402-4896/abfb23} {\bibfield  {journal} {\bibinfo
  {journal} {Physica Scripta}\ }\textbf {\bibinfo {volume} {96}},\ \bibinfo
  {pages} {075704} (\bibinfo {year} {2021})}\BibitemShut {NoStop}%
\bibitem [{\citenamefont {Ma}\ \emph {et~al.}(2022)\citenamefont {Ma},
  \citenamefont {Liu}, \citenamefont {Zhang},\ and\ \citenamefont
  {Wang}}]{Ma2022}%
  \BibitemOpen
  \bibfield  {author} {\bibinfo {author} {\bibfnamefont {T.-X.}\ \bibnamefont
  {Ma}}, \bibinfo {author} {\bibfnamefont {J.}~\bibnamefont {Liu}}, \bibinfo
  {author} {\bibfnamefont {C.}~\bibnamefont {Zhang}},\ and\ \bibinfo {author}
  {\bibfnamefont {Y.-S.}\ \bibnamefont {Wang}},\ }\bibfield  {title} {\bibinfo
  {title} {Topological edge and interface states in phoxonic crystal cavity
  chains},\ }\href {https://doi.org/10.1103/physreva.106.043504} {\bibfield
  {journal} {\bibinfo  {journal} {Physical Review A}\ }\textbf {\bibinfo
  {volume} {106}},\ \bibinfo {pages} {043504} (\bibinfo {year}
  {2022})}\BibitemShut {NoStop}%
\bibitem [{\citenamefont {Lei}\ \emph {et~al.}(2022)\citenamefont {Lei},
  \citenamefont {He}, \citenamefont {Liu}, \citenamefont {Liao},\ and\
  \citenamefont {Yu}}]{Lei2022}%
  \BibitemOpen
  \bibfield  {author} {\bibinfo {author} {\bibfnamefont {L.-L.}\ \bibnamefont
  {Lei}}, \bibinfo {author} {\bibfnamefont {L.-J.}\ \bibnamefont {He}},
  \bibinfo {author} {\bibfnamefont {W.-X.}\ \bibnamefont {Liu}}, \bibinfo
  {author} {\bibfnamefont {Q.-H.}\ \bibnamefont {Liao}},\ and\ \bibinfo
  {author} {\bibfnamefont {T.-B.}\ \bibnamefont {Yu}},\ }\bibfield  {title}
  {\bibinfo {title} {Coexistence of photonic and phononic corner states in a
  second-order topological phoxonic crystal},\ }\href
  {https://doi.org/10.1063/5.0127301} {\bibfield  {journal} {\bibinfo
  {journal} {Applied Physics Letters}\ }\textbf {\bibinfo {volume} {121}},\
  \bibinfo {pages} {193103} (\bibinfo {year} {2022})}\BibitemShut {NoStop}%
\bibitem [{\citenamefont {Lin}\ \emph {et~al.}(2013)\citenamefont {Lin},
  \citenamefont {Lin},\ and\ \citenamefont {Hsu}}]{Lin2013}%
  \BibitemOpen
  \bibfield  {author} {\bibinfo {author} {\bibfnamefont {T.-R.}\ \bibnamefont
  {Lin}}, \bibinfo {author} {\bibfnamefont {C.-H.}\ \bibnamefont {Lin}},\ and\
  \bibinfo {author} {\bibfnamefont {J.-C.}\ \bibnamefont {Hsu}},\ }\bibfield
  {title} {\bibinfo {title} {Enhanced acousto-optic interaction in
  two-dimensional phoxonic crystals with a line defect},\ }\href
  {https://doi.org/10.1063/1.4790288} {\bibfield  {journal} {\bibinfo
  {journal} {Journal of Applied Physics}\ }\textbf {\bibinfo {volume} {113}},\
  \bibinfo {pages} {053508} (\bibinfo {year} {2013})}\BibitemShut {NoStop}%
\bibitem [{\citenamefont {Chiu}\ \emph {et~al.}(2017)\citenamefont {Chiu},
  \citenamefont {Chen}, \citenamefont {Sung},\ and\ \citenamefont
  {Hsiao}}]{Chiu2017}%
  \BibitemOpen
  \bibfield  {author} {\bibinfo {author} {\bibfnamefont {C.-C.}\ \bibnamefont
  {Chiu}}, \bibinfo {author} {\bibfnamefont {W.-M.}\ \bibnamefont {Chen}},
  \bibinfo {author} {\bibfnamefont {K.-W.}\ \bibnamefont {Sung}},\ and\
  \bibinfo {author} {\bibfnamefont {F.-L.}\ \bibnamefont {Hsiao}},\ }\bibfield
  {title} {\bibinfo {title} {High-efficiency acousto-optic coupling in phoxonic
  resonator based on silicon fishbone nanobeam cavity},\ }\href
  {https://doi.org/10.1364/oe.25.006076} {\bibfield  {journal} {\bibinfo
  {journal} {Optics Express}\ }\textbf {\bibinfo {volume} {25}},\ \bibinfo
  {pages} {6076} (\bibinfo {year} {2017})}\BibitemShut {NoStop}%
\bibitem [{\citenamefont {Ma}\ \emph {et~al.}(2014)\citenamefont {Ma},
  \citenamefont {Wang}, \citenamefont {Zhang},\ and\ \citenamefont
  {Su}}]{Ma2014}%
  \BibitemOpen
  \bibfield  {author} {\bibinfo {author} {\bibfnamefont {T.-X.}\ \bibnamefont
  {Ma}}, \bibinfo {author} {\bibfnamefont {Y.-S.}\ \bibnamefont {Wang}},
  \bibinfo {author} {\bibfnamefont {C.}~\bibnamefont {Zhang}},\ and\ \bibinfo
  {author} {\bibfnamefont {X.-X.}\ \bibnamefont {Su}},\ }\bibfield  {title}
  {\bibinfo {title} {Simultaneous guiding of slow elastic and light waves in
  three-dimensional topology-type phoxonic crystals with a line defect},\
  }\href {https://doi.org/10.1088/2040-8978/16/8/085002} {\bibfield  {journal}
  {\bibinfo  {journal} {Journal of Optics}\ }\textbf {\bibinfo {volume} {16}},\
  \bibinfo {pages} {085002} (\bibinfo {year} {2014})}\BibitemShut {NoStop}%
\bibitem [{\citenamefont {Laude}\ \emph {et~al.}(2011)\citenamefont {Laude},
  \citenamefont {Beugnot}, \citenamefont {Benchabane}, \citenamefont {Pennec},
  \citenamefont {Djafari-Rouhani}, \citenamefont {Papanikolaou}, \citenamefont
  {Escalante},\ and\ \citenamefont {Martinez}}]{Laude2011}%
  \BibitemOpen
  \bibfield  {author} {\bibinfo {author} {\bibfnamefont {V.}~\bibnamefont
  {Laude}}, \bibinfo {author} {\bibfnamefont {J.-C.}\ \bibnamefont {Beugnot}},
  \bibinfo {author} {\bibfnamefont {S.}~\bibnamefont {Benchabane}}, \bibinfo
  {author} {\bibfnamefont {Y.}~\bibnamefont {Pennec}}, \bibinfo {author}
  {\bibfnamefont {B.}~\bibnamefont {Djafari-Rouhani}}, \bibinfo {author}
  {\bibfnamefont {N.}~\bibnamefont {Papanikolaou}}, \bibinfo {author}
  {\bibfnamefont {J.~M.}\ \bibnamefont {Escalante}},\ and\ \bibinfo {author}
  {\bibfnamefont {A.}~\bibnamefont {Martinez}},\ }\bibfield  {title} {\bibinfo
  {title} {Simultaneous guidance of slow photons and slow acoustic phonons in
  silicon phoxonic crystal slabs},\ }\href
  {https://doi.org/10.1364/oe.19.009690} {\bibfield  {journal} {\bibinfo
  {journal} {Optics Express}\ }\textbf {\bibinfo {volume} {19}},\ \bibinfo
  {pages} {9690} (\bibinfo {year} {2011})}\BibitemShut {NoStop}%
\bibitem [{\citenamefont {Shu}\ \emph {et~al.}(2020{\natexlab{a}})\citenamefont
  {Shu}, \citenamefont {Yu}, \citenamefont {Yu}, \citenamefont {Liu},
  \citenamefont {Wang},\ and\ \citenamefont {Liao}}]{Shu2020}%
  \BibitemOpen
  \bibfield  {author} {\bibinfo {author} {\bibfnamefont {Y.}~\bibnamefont
  {Shu}}, \bibinfo {author} {\bibfnamefont {M.}~\bibnamefont {Yu}}, \bibinfo
  {author} {\bibfnamefont {T.}~\bibnamefont {Yu}}, \bibinfo {author}
  {\bibfnamefont {W.}~\bibnamefont {Liu}}, \bibinfo {author} {\bibfnamefont
  {T.}~\bibnamefont {Wang}},\ and\ \bibinfo {author} {\bibfnamefont
  {Q.}~\bibnamefont {Liao}},\ }\bibfield  {title} {\bibinfo {title} {Design of
  phoxonic virtual waveguides for both electromagnetic and elastic waves based
  on the self-collimation effect: an application to enhance acousto-optic
  interaction},\ }\href {https://doi.org/10.1364/oe.399591} {\bibfield
  {journal} {\bibinfo  {journal} {Optics Express}\ }\textbf {\bibinfo {volume}
  {28}},\ \bibinfo {pages} {24813} (\bibinfo {year}
  {2020}{\natexlab{a}})}\BibitemShut {NoStop}%
\bibitem [{\citenamefont {Mart{\'{\i}}nez}\ \emph {et~al.}(2022)\citenamefont
  {Mart{\'{\i}}nez}, \citenamefont {Laforge}, \citenamefont {Kadic},\ and\
  \citenamefont {Laude}}]{Martinez2022}%
  \BibitemOpen
  \bibfield  {author} {\bibinfo {author} {\bibfnamefont {J.~A.~I.}\
  \bibnamefont {Mart{\'{\i}}nez}}, \bibinfo {author} {\bibfnamefont
  {N.}~\bibnamefont {Laforge}}, \bibinfo {author} {\bibfnamefont
  {M.}~\bibnamefont {Kadic}},\ and\ \bibinfo {author} {\bibfnamefont
  {V.}~\bibnamefont {Laude}},\ }\bibfield  {title} {\bibinfo {title}
  {Topological waves guided by a glide-reflection symmetric crystal
  interface},\ }\href {https://doi.org/10.1103/physrevb.106.064304} {\bibfield
  {journal} {\bibinfo  {journal} {Physical Review B}\ }\textbf {\bibinfo
  {volume} {106}},\ \bibinfo {pages} {064304} (\bibinfo {year}
  {2022})}\BibitemShut {NoStop}%
\bibitem [{\citenamefont {Tuevedo-Teruel}\ \emph {et~al.}(2021)\citenamefont
  {Tuevedo-Teruel}, \citenamefont {Chen}, \citenamefont {Mesa}, \citenamefont
  {Fonseca},\ and\ \citenamefont {Valerio}}]{QUEVEDOTERUEL2021}%
  \BibitemOpen
  \bibfield  {author} {\bibinfo {author} {\bibfnamefont {O.}~\bibnamefont
  {Tuevedo-Teruel}}, \bibinfo {author} {\bibfnamefont {Q.}~\bibnamefont
  {Chen}}, \bibinfo {author} {\bibfnamefont {F.}~\bibnamefont {Mesa}}, \bibinfo
  {author} {\bibfnamefont {N.~J.~G.}\ \bibnamefont {Fonseca}},\ and\ \bibinfo
  {author} {\bibfnamefont {G.}~\bibnamefont {Valerio}},\ }\bibfield  {title}
  {\bibinfo {title} {On the benefits of glide symmetries for microwave
  devices},\ }\href {https://doi.org/10.1109/jmw.2020.3033847} {\bibfield
  {journal} {\bibinfo  {journal} {{IEEE} Journal of Microwaves}\ }\textbf
  {\bibinfo {volume} {1}},\ \bibinfo {pages} {457} (\bibinfo {year}
  {2021})}\BibitemShut {NoStop}%
\bibitem [{\citenamefont {Dahlberg}\ \emph {et~al.}(2017)\citenamefont
  {Dahlberg}, \citenamefont {Mitchell-Thomas},\ and\ \citenamefont
  {Quevedo-Teruel}}]{Dahlberg2017}%
  \BibitemOpen
  \bibfield  {author} {\bibinfo {author} {\bibfnamefont {O.}~\bibnamefont
  {Dahlberg}}, \bibinfo {author} {\bibfnamefont {R.~C.}\ \bibnamefont
  {Mitchell-Thomas}},\ and\ \bibinfo {author} {\bibfnamefont {O.}~\bibnamefont
  {Quevedo-Teruel}},\ }\bibfield  {title} {\bibinfo {title} {Reducing the
  dispersion of periodic structures with twist and polar glide symmetries},\
  }\bibfield  {journal} {\bibinfo  {journal} {Scientific Reports}\ }\textbf
  {\bibinfo {volume} {7}},\ \href {https://doi.org/10.1038/s41598-017-10566-w}
  {10.1038/s41598-017-10566-w} (\bibinfo {year} {2017})\BibitemShut {NoStop}%
\bibitem [{\citenamefont {Beadle}\ \emph {et~al.}(2019)\citenamefont {Beadle},
  \citenamefont {Hooper}, \citenamefont {Sambles},\ and\ \citenamefont
  {Hibbins}}]{Beadle2019}%
  \BibitemOpen
  \bibfield  {author} {\bibinfo {author} {\bibfnamefont {J.~G.}\ \bibnamefont
  {Beadle}}, \bibinfo {author} {\bibfnamefont {I.~R.}\ \bibnamefont {Hooper}},
  \bibinfo {author} {\bibfnamefont {J.~R.}\ \bibnamefont {Sambles}},\ and\
  \bibinfo {author} {\bibfnamefont {A.~P.}\ \bibnamefont {Hibbins}},\
  }\bibfield  {title} {\bibinfo {title} {Broadband, slow sound on a
  glide-symmetric meander-channel surface},\ }\href
  {https://doi.org/10.1121/1.5109549} {\bibfield  {journal} {\bibinfo
  {journal} {The Journal of the Acoustical Society of America}\ }\textbf
  {\bibinfo {volume} {145}},\ \bibinfo {pages} {3190} (\bibinfo {year}
  {2019})}\BibitemShut {NoStop}%
\bibitem [{\citenamefont {Jankovi{\'{c}}}\ and\ \citenamefont
  {Al{\`{u}}}(2021)}]{Jankovic2021}%
  \BibitemOpen
  \bibfield  {author} {\bibinfo {author} {\bibfnamefont {N.}~\bibnamefont
  {Jankovi{\'{c}}}}\ and\ \bibinfo {author} {\bibfnamefont {A.}~\bibnamefont
  {Al{\`{u}}}},\ }\bibfield  {title} {\bibinfo {title} {Glide-symmetric
  acoustic waveguides for extreme sensing and isolation},\ }\href
  {https://doi.org/10.1103/physrevapplied.15.024004} {\bibfield  {journal}
  {\bibinfo  {journal} {Physical Review Applied}\ }\textbf {\bibinfo {volume}
  {15}},\ \bibinfo {pages} {024004} (\bibinfo {year} {2021})}\BibitemShut
  {NoStop}%
\bibitem [{\citenamefont {Abdollahy}\ \emph {et~al.}(2021)\citenamefont
  {Abdollahy}, \citenamefont {Farahbakhsh},\ and\ \citenamefont
  {Ostovarzadeh}}]{Abdollahy2021}%
  \BibitemOpen
  \bibfield  {author} {\bibinfo {author} {\bibfnamefont {H.}~\bibnamefont
  {Abdollahy}}, \bibinfo {author} {\bibfnamefont {A.}~\bibnamefont
  {Farahbakhsh}},\ and\ \bibinfo {author} {\bibfnamefont {M.~H.}\ \bibnamefont
  {Ostovarzadeh}},\ }\bibfield  {title} {\bibinfo {title} {Mechanical
  reconfigurable phase shifter based on gap waveguide technology},\ }\href
  {https://doi.org/10.1016/j.aeue.2021.153655} {\bibfield  {journal} {\bibinfo
  {journal} {{AEU} - International Journal of Electronics and Communications}\
  }\textbf {\bibinfo {volume} {132}},\ \bibinfo {pages} {153655} (\bibinfo
  {year} {2021})}\BibitemShut {NoStop}%
\bibitem [{\citenamefont {Mahmoodian}\ \emph {et~al.}(2016)\citenamefont
  {Mahmoodian}, \citenamefont {Prindal-Nielsen}, \citenamefont {Söllner},
  \citenamefont {Stobbe},\ and\ \citenamefont {Lodahl}}]{Mahmoodian2016}%
  \BibitemOpen
  \bibfield  {author} {\bibinfo {author} {\bibfnamefont {S.}~\bibnamefont
  {Mahmoodian}}, \bibinfo {author} {\bibfnamefont {K.}~\bibnamefont
  {Prindal-Nielsen}}, \bibinfo {author} {\bibfnamefont {I.}~\bibnamefont
  {Söllner}}, \bibinfo {author} {\bibfnamefont {S.}~\bibnamefont {Stobbe}},\
  and\ \bibinfo {author} {\bibfnamefont {P.}~\bibnamefont {Lodahl}},\
  }\bibfield  {title} {\bibinfo {title} {Engineering chiral
  light{\textendash}matter interaction in photonic crystal waveguides with slow
  light},\ }\href {https://doi.org/10.1364/ome.7.000043} {\bibfield  {journal}
  {\bibinfo  {journal} {Optical Materials Express}\ }\textbf {\bibinfo {volume}
  {7}},\ \bibinfo {pages} {43} (\bibinfo {year} {2016})}\BibitemShut {NoStop}%
\bibitem [{\citenamefont {Yoshimi}\ \emph {et~al.}(2020)\citenamefont
  {Yoshimi}, \citenamefont {Yamaguchi}, \citenamefont {Ota}, \citenamefont
  {Arakawa},\ and\ \citenamefont {Iwamoto}}]{Yoshimi2020}%
  \BibitemOpen
  \bibfield  {author} {\bibinfo {author} {\bibfnamefont {H.}~\bibnamefont
  {Yoshimi}}, \bibinfo {author} {\bibfnamefont {T.}~\bibnamefont {Yamaguchi}},
  \bibinfo {author} {\bibfnamefont {Y.}~\bibnamefont {Ota}}, \bibinfo {author}
  {\bibfnamefont {Y.}~\bibnamefont {Arakawa}},\ and\ \bibinfo {author}
  {\bibfnamefont {S.}~\bibnamefont {Iwamoto}},\ }\bibfield  {title} {\bibinfo
  {title} {Slow light waveguides in topological valley photonic crystals},\
  }\href {https://doi.org/10.1364/ol.391764} {\bibfield  {journal} {\bibinfo
  {journal} {Optics Letters}\ }\textbf {\bibinfo {volume} {45}},\ \bibinfo
  {pages} {2648} (\bibinfo {year} {2020})}\BibitemShut {NoStop}%
\bibitem [{\citenamefont {Yoshimi}\ \emph {et~al.}(2021)\citenamefont
  {Yoshimi}, \citenamefont {Yamaguchi}, \citenamefont {Katsumi}, \citenamefont
  {Ota}, \citenamefont {Arakawa},\ and\ \citenamefont {Iwamoto}}]{Yoshimi2021}%
  \BibitemOpen
  \bibfield  {author} {\bibinfo {author} {\bibfnamefont {H.}~\bibnamefont
  {Yoshimi}}, \bibinfo {author} {\bibfnamefont {T.}~\bibnamefont {Yamaguchi}},
  \bibinfo {author} {\bibfnamefont {R.}~\bibnamefont {Katsumi}}, \bibinfo
  {author} {\bibfnamefont {Y.}~\bibnamefont {Ota}}, \bibinfo {author}
  {\bibfnamefont {Y.}~\bibnamefont {Arakawa}},\ and\ \bibinfo {author}
  {\bibfnamefont {S.}~\bibnamefont {Iwamoto}},\ }\bibfield  {title} {\bibinfo
  {title} {Experimental demonstration of topological slow light waveguides in
  valley photonic crystals},\ }\href {https://doi.org/10.1364/oe.422962}
  {\bibfield  {journal} {\bibinfo  {journal} {Optics Express}\ }\textbf
  {\bibinfo {volume} {29}},\ \bibinfo {pages} {13441} (\bibinfo {year}
  {2021})}\BibitemShut {NoStop}%
\bibitem [{\citenamefont {Xie}\ \emph {et~al.}(2021)\citenamefont {Xie},
  \citenamefont {Yan}, \citenamefont {Dang}, \citenamefont {Yang},
  \citenamefont {Xiao}, \citenamefont {Wang}, \citenamefont {Shi},
  \citenamefont {Yang}, \citenamefont {Dai}, \citenamefont {Yuan},
  \citenamefont {Luo}, \citenamefont {Cui}, \citenamefont {Chi}, \citenamefont
  {Zuo}, \citenamefont {Li}, \citenamefont {Wang},\ and\ \citenamefont
  {Xu}}]{Xie2021}%
  \BibitemOpen
  \bibfield  {author} {\bibinfo {author} {\bibfnamefont {X.}~\bibnamefont
  {Xie}}, \bibinfo {author} {\bibfnamefont {S.}~\bibnamefont {Yan}}, \bibinfo
  {author} {\bibfnamefont {J.}~\bibnamefont {Dang}}, \bibinfo {author}
  {\bibfnamefont {J.}~\bibnamefont {Yang}}, \bibinfo {author} {\bibfnamefont
  {S.}~\bibnamefont {Xiao}}, \bibinfo {author} {\bibfnamefont {Y.}~\bibnamefont
  {Wang}}, \bibinfo {author} {\bibfnamefont {S.}~\bibnamefont {Shi}}, \bibinfo
  {author} {\bibfnamefont {L.}~\bibnamefont {Yang}}, \bibinfo {author}
  {\bibfnamefont {D.}~\bibnamefont {Dai}}, \bibinfo {author} {\bibfnamefont
  {Y.}~\bibnamefont {Yuan}}, \bibinfo {author} {\bibfnamefont {N.}~\bibnamefont
  {Luo}}, \bibinfo {author} {\bibfnamefont {T.}~\bibnamefont {Cui}}, \bibinfo
  {author} {\bibfnamefont {G.}~\bibnamefont {Chi}}, \bibinfo {author}
  {\bibfnamefont {Z.}~\bibnamefont {Zuo}}, \bibinfo {author} {\bibfnamefont
  {B.-B.}\ \bibnamefont {Li}}, \bibinfo {author} {\bibfnamefont
  {C.}~\bibnamefont {Wang}},\ and\ \bibinfo {author} {\bibfnamefont
  {X.}~\bibnamefont {Xu}},\ }\bibfield  {title} {\bibinfo {title} {Topological
  cavity based on slow-light topological edge mode for broadband purcell
  enhancement},\ }\href {https://doi.org/10.1103/physrevapplied.16.014036}
  {\bibfield  {journal} {\bibinfo  {journal} {Physical Review Applied}\
  }\textbf {\bibinfo {volume} {16}},\ \bibinfo {pages} {014036} (\bibinfo
  {year} {2021})}\BibitemShut {NoStop}%
\bibitem [{\citenamefont {Wen}\ \emph {et~al.}(2022)\citenamefont {Wen},
  \citenamefont {Bisharat}, \citenamefont {Davis}, \citenamefont {Yang},\ and\
  \citenamefont {Sievenpiper}}]{Wen2022}%
  \BibitemOpen
  \bibfield  {author} {\bibinfo {author} {\bibfnamefont {E.}~\bibnamefont
  {Wen}}, \bibinfo {author} {\bibfnamefont {D.~J.}\ \bibnamefont {Bisharat}},
  \bibinfo {author} {\bibfnamefont {R.~J.}\ \bibnamefont {Davis}}, \bibinfo
  {author} {\bibfnamefont {X.}~\bibnamefont {Yang}},\ and\ \bibinfo {author}
  {\bibfnamefont {D.~F.}\ \bibnamefont {Sievenpiper}},\ }\bibfield  {title}
  {\bibinfo {title} {Designing topological defect lines protected by
  gauge-dependent symmetry indicators},\ }\href
  {https://doi.org/10.1103/physrevapplied.17.064008} {\bibfield  {journal}
  {\bibinfo  {journal} {Physical Review Applied}\ }\textbf {\bibinfo {volume}
  {17}},\ \bibinfo {pages} {064008} (\bibinfo {year} {2022})}\BibitemShut
  {NoStop}%
\bibitem [{\citenamefont {Mock}\ \emph {et~al.}(2010)\citenamefont {Mock},
  \citenamefont {Lu},\ and\ \citenamefont {O'Brien}}]{Mock2010}%
  \BibitemOpen
  \bibfield  {author} {\bibinfo {author} {\bibfnamefont {A.}~\bibnamefont
  {Mock}}, \bibinfo {author} {\bibfnamefont {L.}~\bibnamefont {Lu}},\ and\
  \bibinfo {author} {\bibfnamefont {J.}~\bibnamefont {O'Brien}},\ }\bibfield
  {title} {\bibinfo {title} {Space group theory and fourier space analysis of
  two-dimensional photonic crystal waveguides},\ }\href
  {https://doi.org/10.1103/physrevb.81.155115} {\bibfield  {journal} {\bibinfo
  {journal} {Physical Review B}\ }\textbf {\bibinfo {volume} {81}},\ \bibinfo
  {pages} {155115} (\bibinfo {year} {2010})}\BibitemShut {NoStop}%
\bibitem [{\citenamefont {Kim}\ and\ \citenamefont {Murakami}(2020)}]{Kim2020}%
  \BibitemOpen
  \bibfield  {author} {\bibinfo {author} {\bibfnamefont {H.}~\bibnamefont
  {Kim}}\ and\ \bibinfo {author} {\bibfnamefont {S.}~\bibnamefont {Murakami}},\
  }\bibfield  {title} {\bibinfo {title} {Glide-symmetric topological
  crystalline insulator phase in a nonprimitive lattice},\ }\href
  {https://doi.org/10.1103/physrevb.102.195202} {\bibfield  {journal} {\bibinfo
   {journal} {Physical Review B}\ }\textbf {\bibinfo {volume} {102}},\ \bibinfo
  {pages} {195202} (\bibinfo {year} {2020})}\BibitemShut {NoStop}%
\bibitem [{\citenamefont {Ghasemifard}\ \emph {et~al.}(2018)\citenamefont
  {Ghasemifard}, \citenamefont {Norgren},\ and\ \citenamefont
  {Quevedo-Teruel}}]{Ghasemifard2018}%
  \BibitemOpen
  \bibfield  {author} {\bibinfo {author} {\bibfnamefont {F.}~\bibnamefont
  {Ghasemifard}}, \bibinfo {author} {\bibfnamefont {M.}~\bibnamefont
  {Norgren}},\ and\ \bibinfo {author} {\bibfnamefont {O.}~\bibnamefont
  {Quevedo-Teruel}},\ }\bibfield  {title} {\bibinfo {title} {Twist and polar
  glide symmetries: an additional degree of freedom to control the propagation
  characteristics of periodic structures},\ }\bibfield  {journal} {\bibinfo
  {journal} {Scientific Reports}\ }\textbf {\bibinfo {volume} {8}},\ \href
  {https://doi.org/10.1038/s41598-018-29565-6} {10.1038/s41598-018-29565-6}
  (\bibinfo {year} {2018})\BibitemShut {NoStop}%
\bibitem [{\citenamefont {Zhang}\ and\ \citenamefont {Zhou}(2017)}]{Zhang2017}%
  \BibitemOpen
  \bibfield  {author} {\bibinfo {author} {\bibfnamefont {S.-L.}\ \bibnamefont
  {Zhang}}\ and\ \bibinfo {author} {\bibfnamefont {Q.}~\bibnamefont {Zhou}},\
  }\bibfield  {title} {\bibinfo {title} {Two-leg su-schrieffer-heeger chain
  with glide reflection symmetry},\ }\href
  {https://doi.org/10.1103/physreva.95.061601} {\bibfield  {journal} {\bibinfo
  {journal} {Physical Review A}\ }\textbf {\bibinfo {volume} {95}},\ \bibinfo
  {pages} {061601} (\bibinfo {year} {2017})}\BibitemShut {NoStop}%
\bibitem [{\citenamefont {Nica}\ \emph {et~al.}(2015)\citenamefont {Nica},
  \citenamefont {Yu},\ and\ \citenamefont {Si}}]{Nica2015}%
  \BibitemOpen
  \bibfield  {author} {\bibinfo {author} {\bibfnamefont {E.~M.}\ \bibnamefont
  {Nica}}, \bibinfo {author} {\bibfnamefont {R.}~\bibnamefont {Yu}},\ and\
  \bibinfo {author} {\bibfnamefont {Q.}~\bibnamefont {Si}},\ }\bibfield
  {title} {\bibinfo {title} {Glide reflection symmetry, brillouin zone folding,
  and superconducting pairing for the p4/nmm space group},\ }\href
  {https://doi.org/10.1103/physrevb.92.174520} {\bibfield  {journal} {\bibinfo
  {journal} {Physical Review B}\ }\textbf {\bibinfo {volume} {92}},\ \bibinfo
  {pages} {174520} (\bibinfo {year} {2015})}\BibitemShut {NoStop}%
\bibitem [{\citenamefont {Lei}\ \emph {et~al.}(2021)\citenamefont {Lei},
  \citenamefont {Yu}, \citenamefont {Liu}, \citenamefont {Wang},\ and\
  \citenamefont {Liao}}]{Lei2021}%
  \BibitemOpen
  \bibfield  {author} {\bibinfo {author} {\bibfnamefont {L.}~\bibnamefont
  {Lei}}, \bibinfo {author} {\bibfnamefont {T.}~\bibnamefont {Yu}}, \bibinfo
  {author} {\bibfnamefont {W.}~\bibnamefont {Liu}}, \bibinfo {author}
  {\bibfnamefont {T.}~\bibnamefont {Wang}},\ and\ \bibinfo {author}
  {\bibfnamefont {Q.}~\bibnamefont {Liao}},\ }\bibfield  {title} {\bibinfo
  {title} {Dirac cones with zero refractive indices in phoxonic crystals},\
  }\href {https://doi.org/10.1364/oe.446356} {\bibfield  {journal} {\bibinfo
  {journal} {Optics Express}\ }\textbf {\bibinfo {volume} {30}},\ \bibinfo
  {pages} {308} (\bibinfo {year} {2021})}\BibitemShut {NoStop}%
\bibitem [{\citenamefont {Shu}\ \emph {et~al.}(2020{\natexlab{b}})\citenamefont
  {Shu}, \citenamefont {Yu}, \citenamefont {Yu}, \citenamefont {Liu},
  \citenamefont {Wang},\ and\ \citenamefont {Liao}}]{Shu2020a}%
  \BibitemOpen
  \bibfield  {author} {\bibinfo {author} {\bibfnamefont {Y.}~\bibnamefont
  {Shu}}, \bibinfo {author} {\bibfnamefont {M.}~\bibnamefont {Yu}}, \bibinfo
  {author} {\bibfnamefont {T.}~\bibnamefont {Yu}}, \bibinfo {author}
  {\bibfnamefont {W.}~\bibnamefont {Liu}}, \bibinfo {author} {\bibfnamefont
  {T.}~\bibnamefont {Wang}},\ and\ \bibinfo {author} {\bibfnamefont
  {Q.}~\bibnamefont {Liao}},\ }\bibfield  {title} {\bibinfo {title} {Design of
  phoxonic virtual waveguides for both electromagnetic and elastic waves based
  on the self-collimation effect: an application to enhance acousto-optic
  interaction},\ }\href {https://doi.org/10.1364/OE.399591} {\bibfield
  {journal} {\bibinfo  {journal} {Opt. Express}\ }\textbf {\bibinfo {volume}
  {28}},\ \bibinfo {pages} {24813} (\bibinfo {year}
  {2020}{\natexlab{b}})}\BibitemShut {NoStop}%
\bibitem [{\citenamefont {Lei}\ \emph {et~al.}(2023)\citenamefont {Lei},
  \citenamefont {Xiao}, \citenamefont {Liu}, \citenamefont {Liao},
  \citenamefont {He},\ and\ \citenamefont {Yu}}]{Lei2023}%
  \BibitemOpen
  \bibfield  {author} {\bibinfo {author} {\bibfnamefont {L.}~\bibnamefont
  {Lei}}, \bibinfo {author} {\bibfnamefont {S.}~\bibnamefont {Xiao}}, \bibinfo
  {author} {\bibfnamefont {W.}~\bibnamefont {Liu}}, \bibinfo {author}
  {\bibfnamefont {Q.}~\bibnamefont {Liao}}, \bibinfo {author} {\bibfnamefont
  {L.}~\bibnamefont {He}},\ and\ \bibinfo {author} {\bibfnamefont
  {T.}~\bibnamefont {Yu}},\ }\bibfield  {title} {\bibinfo {title}
  {Polarization-independent second-order photonic topological corner states},\
  }\href {https://doi.org/10.1103/PhysRevApplied.20.024014} {\bibfield
  {journal} {\bibinfo  {journal} {Phys. Rev. Appl.}\ }\textbf {\bibinfo
  {volume} {20}},\ \bibinfo {pages} {024014} (\bibinfo {year}
  {2023})}\BibitemShut {NoStop}%
\bibitem [{\citenamefont {Yu}\ and\ \citenamefont {Sun}(2018)}]{Yu2018}%
  \BibitemOpen
  \bibfield  {author} {\bibinfo {author} {\bibfnamefont {Z.}~\bibnamefont
  {Yu}}\ and\ \bibinfo {author} {\bibfnamefont {X.}~\bibnamefont {Sun}},\
  }\bibfield  {title} {\bibinfo {title} {Giant enhancement of stimulated
  brillouin scattering with engineered phoxonic crystal waveguides},\ }\href
  {https://doi.org/10.1364/oe.26.001255} {\bibfield  {journal} {\bibinfo
  {journal} {Optics Express}\ }\textbf {\bibinfo {volume} {26}},\ \bibinfo
  {pages} {1255} (\bibinfo {year} {2018})}\BibitemShut {NoStop}%
\bibitem [{\citenamefont {Lin}\ \emph {et~al.}(2020)\citenamefont {Lin},
  \citenamefont {Wang}, \citenamefont {Xiong}, \citenamefont {Lu},\ and\
  \citenamefont {Jiang}}]{Lin2020}%
  \BibitemOpen
  \bibfield  {author} {\bibinfo {author} {\bibfnamefont {Z.-K.}\ \bibnamefont
  {Lin}}, \bibinfo {author} {\bibfnamefont {H.-X.}\ \bibnamefont {Wang}},
  \bibinfo {author} {\bibfnamefont {Z.}~\bibnamefont {Xiong}}, \bibinfo
  {author} {\bibfnamefont {M.-H.}\ \bibnamefont {Lu}},\ and\ \bibinfo {author}
  {\bibfnamefont {J.-H.}\ \bibnamefont {Jiang}},\ }\bibfield  {title} {\bibinfo
  {title} {Anomalous quadrupole topological insulators in two-dimensional
  nonsymmorphic sonic crystals},\ }\href
  {https://doi.org/10.1103/physrevb.102.035105} {\bibfield  {journal} {\bibinfo
   {journal} {Physical Review B}\ }\textbf {\bibinfo {volume} {102}},\ \bibinfo
  {pages} {035105} (\bibinfo {year} {2020})}\BibitemShut {NoStop}%
\bibitem [{\citenamefont {El-jallal}\ \emph {et~al.}(2013)\citenamefont
  {El-jallal}, \citenamefont {Oudich}, \citenamefont {Pennec}, \citenamefont
  {Djafari-Rouhani}, \citenamefont {Makhoute}, \citenamefont {Rolland},
  \citenamefont {Dupont},\ and\ \citenamefont {Gazalet}}]{Eljallal2013}%
  \BibitemOpen
  \bibfield  {author} {\bibinfo {author} {\bibfnamefont {S.}~\bibnamefont
  {El-jallal}}, \bibinfo {author} {\bibfnamefont {M.}~\bibnamefont {Oudich}},
  \bibinfo {author} {\bibfnamefont {Y.}~\bibnamefont {Pennec}}, \bibinfo
  {author} {\bibfnamefont {B.}~\bibnamefont {Djafari-Rouhani}}, \bibinfo
  {author} {\bibfnamefont {A.}~\bibnamefont {Makhoute}}, \bibinfo {author}
  {\bibfnamefont {Q.}~\bibnamefont {Rolland}}, \bibinfo {author} {\bibfnamefont
  {S.}~\bibnamefont {Dupont}},\ and\ \bibinfo {author} {\bibfnamefont
  {J.}~\bibnamefont {Gazalet}},\ }\bibfield  {title} {\bibinfo {title}
  {Optomechanical interactions in two-dimensional si and gaas phoxonic
  cavities},\ }\href {https://doi.org/10.1088/0953-8984/26/1/015005} {\bibfield
   {journal} {\bibinfo  {journal} {Journal of Physics: Condensed Matter}\
  }\textbf {\bibinfo {volume} {26}},\ \bibinfo {pages} {015005} (\bibinfo
  {year} {2013})}\BibitemShut {NoStop}%
\end{thebibliography}%

\end{document}